\documentclass[journal=jctcce,manuscript=article,layout=twocolumn]{achemso}

\usepackage[colorlinks,linkcolor=blue,citecolor=blue,urlcolor=black,bookmarks=false,hypertexnames=true]{hyperref} 

\usepackage[version=3]{mhchem}
\usepackage{amssymb}
\usepackage{color}
\usepackage{braket}
\usepackage{xspace}
\usepackage{cleveref}
\usepackage{graphicx}
\usepackage{subfigure}
\usepackage{threeparttable}
\usepackage{textcomp}
\usepackage{setspace}
\usepackage{caption}
\usepackage{comment}

\usepackage{array}
\newcolumntype{L}[1]{>{\raggedright\let\newline\\\arraybackslash\hspace{0pt}}m{#1}}
\newcolumntype{C}[1]{>{\centering\let\newline\\\arraybackslash\hspace{0pt}}m{#1}}
\newcolumntype{R}[1]{>{\raggedleft\let\newline\\\arraybackslash\hspace{0pt}}m{#1}}

\makeatletter
\let\l@addto@macro\relax
\makeatother
\usepackage[fontsize=11pt]{scrextend}

\let\oldmaketitle\maketitle
\let\maketitle\relax

\newcommand{\h}[2]{h_{{#1}}^{{#2}}}
\newcommand{\f}[2]{f_{{#1}}^{{#2}}}

\renewcommand{\v}[2]{{v}_{{#1}}^{{#2}}}

\renewcommand{\c}[1]{a^\dagger_{#1}}
\renewcommand{\a}[1]{a_{#1}}

\newcommand{\cm}{\ensuremath{\text{cm}^{-1}}\xspace}

\newcommand{\angstrom}{\mbox{\normalfont\AA}\xspace}

\DeclareUnicodeCharacter{2009}{\,}

\crefname{figure}{Figure}{Figures}
\crefname{table}{Table}{Tables}
\crefname{equation}{Eq.}{Eqs.}
\crefname{section}{Section}{Sections}
\crefname{subsection}{Section}{Sections}

\author{Rajat~Majumder$^\dag$}
\affiliation{$^\dag$Department of Chemistry and Biochemistry, The Ohio State University, Columbus, Ohio 43210, USA}

\author{Alexander~Yu.~Sokolov$^\dag$}
\email{sokolov.8@osu.edu}
\affiliation{$^\dag$Department of Chemistry and Biochemistry, The Ohio State University, Columbus, Ohio 43210, USA}

\begin{tocentry}
\includegraphics[width=1.0\textwidth]{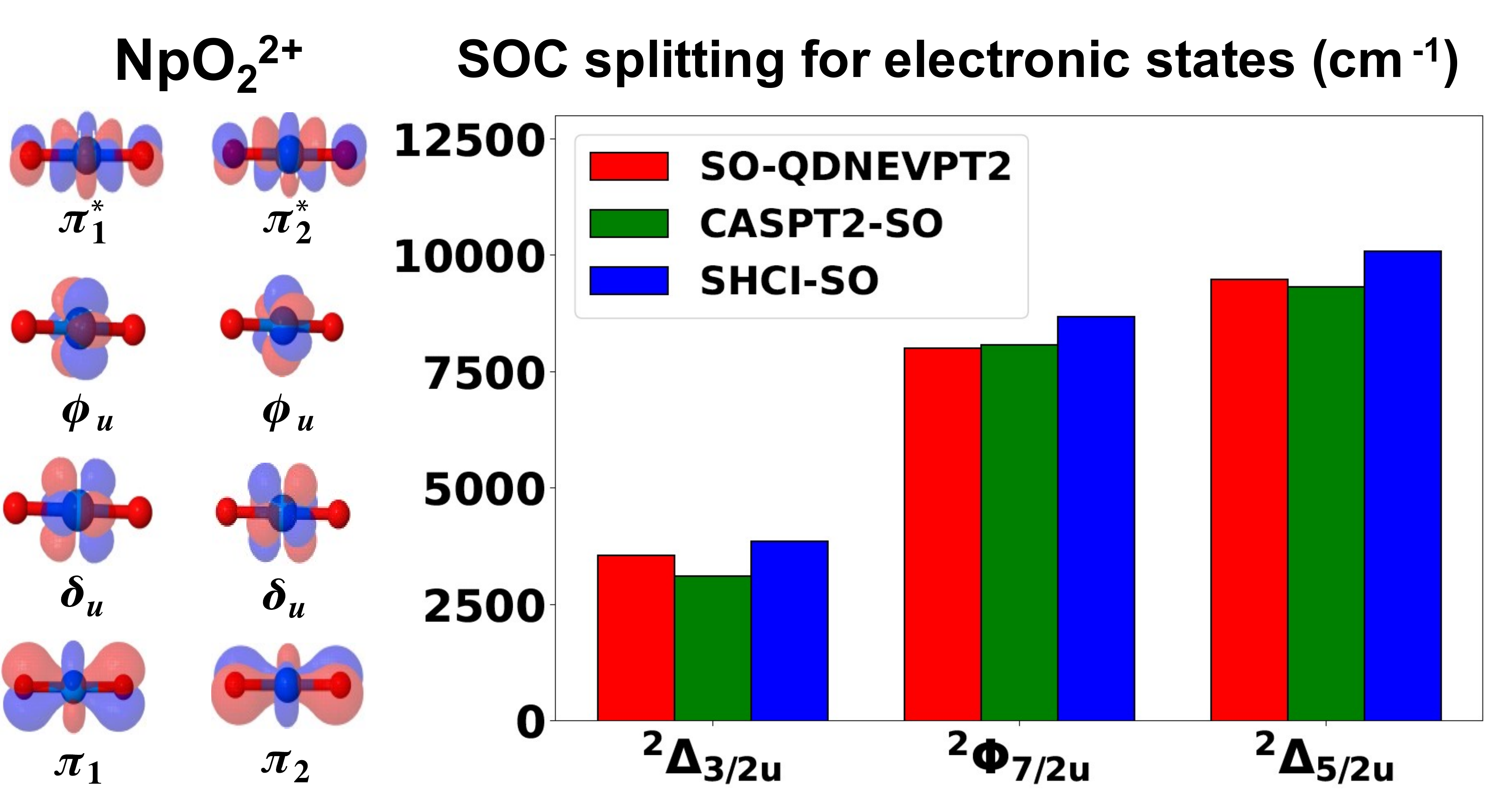}
\end{tocentry}

\title{{\color{blue}Simulating Spin--Orbit Coupling With Quasidegenerate $N$-Electron Valence Perturbation Theory}}

\begin{document}

\setstretch{1.0}

\newcommand*{\abstractext}{
We present the first implementation of spin--orbit coupling effects in fully internally contracted second-order quasidegenerate $N$-electron valence perturbation theory (SO-QDNEVPT2).
The SO-QDNEVPT2 approach enables the computations of ground- and excited-state energies and oscillator strengths combining the description of static electron correlation with an efficient treatment of dynamic correlation and spin--orbit coupling.
In addition to SO-QDNEVPT2 with the full description of one- and two-body spin--orbit interactions at the level of two-component Breit--Pauli Hamiltonian, our implementation also features a simplified approach that takes advantage of spin--orbit mean-field approximation (SOMF-QDNEVPT2).
The accuracy of these methods is tested for the group 14 and 16 hydrides, $3d$ and $4d$ transition metal ions, and two actinide dioxides (neptunyl and plutonyl dications). 
The zero-field splittings of group 14 and 16 molecules computed using SO-QDNEVPT2 and SOMF-QDNEVPT2 are in a good agreement with the available experimental data.
For the $3d$ transition metal ions, the SO-QDNEVPT2 method is significantly more accurate than SOMF-QDNEVPT2, while no substantial difference in the performance of two methods is observed for the $4d$ ions.
Finally, we demonstrate that for the actinide dioxides the results of SO-QDNEVPT2 and SOMF-QDNEVPT2 are in a good agreement with the data from previous theoretical studies of these systems.
Overall, our results demonstrate that SO-QDNEVPT2 and SOMF-QDNEVPT2 are promising multireference methods for treating spin--orbit coupling with a relatively low computational cost.
\vspace{0.25cm}
}

\twocolumn[
\begin{@twocolumnfalse}
\oldmaketitle
\vspace{-0.75cm}
\begin{abstract}
\abstractext
\end{abstract}
\end{@twocolumnfalse}
]

\section{Introduction}
\label{sec:introduction}

Relativistic effects play a major role in how molecules and materials interact with light.
Among different types of relativistic interactions, spin--orbit coupling is of particular importance, giving rise to a variety of experimentally observed phenomena, such as zero-field splitting, intersystem crossing, and magnetism.\cite{Pyykko:2012p45}
Spin--orbit coupling becomes increasingly significant in the ground and low-lying excited states of elements starting with the fourth row of periodic table and has a profound influence on the electronic structure of compounds with heavier elements ($>$ 5th row).\cite{Cao:2017p3713,Malmqvist:2002p230}
For the lighter elements, spin--orbit coupling is important in the core-level excited states that can be accessed by the excitations with X-ray radiation.\cite{Lee:2010p9715,Kasper:2018p1998,Maganas:2019p104106,Carbone:2019p241,Stetina:2019p234103,Vidal:2020p8314} 

Detailed understanding of spin--orbit-coupled states requires insights from accurate relativistic electronic structure calculations.
However, incorporating spin--orbit coupling into the simulations of light--matter interactions introduces new challenges for electronic structure theories.
These challenges include using a more complicated relativistic Hamiltonian, treating the coupling between electronic and positronic states in the Dirac equation, and employing large (uncontracted or reparametrized) basis sets.\cite{Dyall:1995_book,Saue:2011p3077,Reiher:2014_book}
For this reason, relativistic electronic structure methods \cite{ Douglas:1974p89,VanLenthe:1993p4597,Barysz:1997p225239,Sadlej:1998p1758,Dyall:1998p9618,Neese:1998p6568,Wolf:2002p9215,Barysz:2002p2696,Reiher:2004p2037,Reiher:2004p10945,Neese:2005p034107,Kutzelnigg:2005p241102,Ganyushin:2006p024103,Liu:2006p044102,Ilias:2007p064102,Peng:2007p104106,Liu:2009p031104,Kutzelnigg:2012p16,Peng:2013p184105,Cheng:2014p164107,Epifanovsky:2015p64102,Egidi:2016p3711,Konecny:2016p5823,Meitei:2020p3597} have a higher computational cost than their nonrelativistic counterparts, which limits their applications to smaller chemical systems.
In practical calculations, the description of spin--orbit coupling must be combined with an accurate treatment of electron--electron interactions, ranging from static electron correlation in valence molecular orbitals to dynamic correlation of inner-shell and core electrons.

An attractive approach for treating electron correlation in molecules is quasidegenerate second-order $N$-electron valence perturbation theory (QDNEVPT2).\cite{Angeli:2001p10252,Angeli:2004p4043} 
QDNEVPT2 is an intruder-free multistate multireference perturbation theory, which enables an accurate treatment of static and dynamic correlation in near-degenerate electronic states with a relatively low computational cost.
Several implementations of QDNEVPT2 that are different in the degree of internal contraction in multireference wavefunctions have been developed, namely: i) strongly contracted (sc-QDNEVPT2),\cite{Angeli:2004p4043} ii) partially or fully internally contracted (pc-QDNEVPT2),\cite{Angeli:2004p4043,Park:2019p5417,Nishimoto:2020p137219} and iii) uncontracted (uc-QDNEVPT2).\cite{Sharma:2016p034103}
Out of these three variants, only sc-QDNEVPT2 has been extended to incorporate spin--orbit coupling effects and calculate zero-field splitting parameters\cite{Neese:2020p224108} within the formalism of spin--orbit mean-field (SOMF) approximation.\cite{Hess:1996p365,Berning:2000p1823}
In addition, spin--orbit coupling has been implemented in strongly and fully internally contracted state-specific NEVPT2 (sc- and pc-NEVPT2).\cite{Ganyushin:2006p024103,Neese:2007p164112,Duboc:2010p10750,Maurice:2011p6229,Atanasov:2012p12324,Atanasov:2015p177}
Although these methods have been applied to a variety of chemical systems,\cite{Maurice:2011p6229,Atanasov:2012p12324,Retegan:2014p11785,Atanasov:2015p177,Lang:2020p1025} strong contraction in sc-QDNEVPT2 introduces significant errors in correlation energy and violates orbital invariance, leading to numerical instabilities in the evaluation of excited-state properties and optimization of molecular geometries.\cite{Guo:2016p094111,Sokolov:2016p064102,Sivalingam:2016p054104,Park:2019p5417}
Meanwhile, the state-specific sc- and pc-NEVPT2 methods do not correctly describe the interaction between nearly degenerate electronic states, which is particularly important when spin--orbit coupling is taken into account.

Here, we present the first implementation of pc-QDNEVPT2 that combines a computationally efficient description of spin--orbit coupling and electron correlation in the ground and excited electronic states. 
Compared to earlier work, our implementation of pc-QDNEVPT2 has a number of important advantages: i) it avoids the orbital invariance problems inherent in sc-QDNEVPT2 and correctly treats the interaction between nearly degenerate spin--orbit-coupled electronic states that is missing in state-specific theories; ii) it enables the calculations with and without the SOMF approximation, thus allowing to quantify its errors; iii) it does not require calculating the four-particle reduced density matrices, significantly lowering the computational cost; iv) it preserves the degeneracy of electronic states that could otherwise be lost when introducing internal contraction; and v) it allows to calculate excited-state and transition properties, such as oscillator strengths. 

This paper is organized as follows. 
First, we briefly review the theoretical background behind pc-QDNEVPT2 and describe its formulation that incorporates spin--orbit coupling (\cref{sec:theory}).
Next, having discussed the details of our implementation and computations (\cref{sec:implementation,sec:comp_details}), we use pc-QDNEVPT2 to calculate the zero-field splitting in group 14 and 16 hydrides, the spin--orbit coupling constants of $3d$ and $4d$ transition metal ions, and the excited-state energies of neptunyl and plutonyl oxides (\ce{NpO2^2+} and \ce{PuO2^2+}, \cref{sec:results}).
We summarize all findings of this work and outline directions for future developments in \cref{sec:conclusions} .

\section{Theory}
\label{sec:theory}

\subsection{Overview of $N$-electron valence perturbation theory}
\label{sec:theory:nevpt}

Let us consider an $N$-electron system described by a nonrelativistic Hamiltonian $\hat{\cal{H}}$. 
Introducing a finite basis of spin-orbitals $\{ \psi_p \}$, the Hamiltonian $\hat{\cal{H}}$ can be expressed, in second quantization, as:
\begin{align}
	\label{eq:hamiltonian}
	\hat{\cal{H}} = \sum_{pq} \h{p}{q} \c{p}\a{q} + \frac{1}{4} \sum_{pqrs} \v{pq}{rs} \c{p}\c{q}\a{s}\a{r} \ ,
\end{align}
where $ \h{p}{q}$ and $\v{pq}{rs}$ are the one- and antisymmetrized two-electron integrals.
The operators $\c{p}$ and $\a{p}$ create or annihilate a particle, respectively, in a spin-orbital $\psi_p$. 
To describe electron correlation in this system, we partition all spin-orbitals into three subsets, namely: {\it core} (doubly occupied) with indices $i$, $j$, $k$, $l$; {\it active} (usually, frontier) with indices $u$, $v$, $w$, $x$, $y$, $z$; and {\it external} (unoccupied) with indices $a$, $b$, $c$, $d$.

In $N$-electron valence perturbation theory (NEVPT),\cite{Angeli:2001p10252,Angeli:2002p9138,Angeli:2006p054108} the correlation in active orbitals is described by constructing a complete active-space (CAS) wavefunction\cite{Hinze:1973p6424,Werner:1980p5794,Roos:1980p157,Werner:1985p5053,Siegbahn:1998p2384} $\ket{\Psi_I^{(0)}}$ for the $I$th electronic state of interest. 
The electron correlation in remaining orbitals (core and external) is incorporated perturbatively by partitioning the Hamiltonian $\hat{\cal{H}}$ into two contributions: the zeroth-order Dyall Hamiltonian\cite{Dyall:1998p4909} 
\begin{align}
	\label{eq:dyall_h}
		\hat{\cal{H}}^{(0)} = C + \sum_{i}\epsilon_{i}a^{\dagger}_{i}a_{i} + \sum_{a}\epsilon_{a}a^{\dagger}_{a}a_{a} +  \hat{\cal{H}}_{active}
\end{align}
and the perturbation operator
\begin{align}
	\hat{\cal{V}} = \hat{\cal{H}} - \hat{\cal{H}}^{(0)} \ .
\end{align}
The Dyall Hamiltonian $\hat{\cal{H}}^{(0)}$ depends on the core ($\epsilon_{i}$) and external ($\epsilon_{a}$) eigenvalues of the generalized Fock matrix
\begin{align}
	\label{eq:gen_fock_rdm}
	f^{q}_{p} = h^{q}_{p} + \sum_{rs}v^{qs}_{pr}\gamma^{r}_{s} \ , \quad 
	\gamma^{q}_{p} = \langle{\Psi_I}|a^{\dagger}_{p}a_{q}|\Psi_I\rangle	 \ ,
\end{align}
the constant term
\begin{align}
	C &= \sum_i \h{i}{i} + \frac{1}{2}\sum_{ij}\v{ij}{ij} - \sum_i  \f{i}{i}  \ ,
\end{align}
and all one- and two-electron terms of the full Hamiltonian in the active space
\begin{align}
	\label{eq:dyall_h_act}
	\hat{\cal{H}}_{active} 
	&=  \sum_{xy} \left( h^{y}_{x} + \sum_{i}v^{yi}_{xi}  \right) a^{\dagger}_{x}a_{y} \notag \\
	&+ \frac{1}{4}\sum_{wxyz}v^{zw}_{xy}a^{\dagger}_{x}a^{\dagger}_{y}a_{w}a_{z} \ ,
\end{align}

Expanding the energy of the $I$th state $E_I = \braket{\Psi_I|\hat{\cal{H}}|\Psi_I}$ with respect to the perturbation $\hat{\cal{V}}$ and truncating the expansion at second order, we obtain the correlation energy of {\it fully uncontracted} second-order $N$-electron valence perturbation theory (uc-NEVPT2):
\begin{align}
	\label{eq:corr_energy}
	E_I^{(2)} 
	&= \braket{\Psi_I^{(0)}|  \hat{\cal{V}}^\dag  \frac{1}{E_I^{(0)} - \hat{\cal{H}}^{(0)}}  \hat{\cal{V}}  |\Psi_I^{(0)}} \notag \\
	&\equiv \braket{\Psi_I^{(0)}|  \hat{\cal{V}}^\dag  |\Psi^{(1)}_I} \ .
\end{align}
\cref{eq:corr_energy} can be evaluated exactly, but requires expanding the first-order wavefunction $\ket{\Psi^{(1)}_I}$ in a very large set of determinants that comprise the first-order interacting space.
As a result, calculating the uc-NEVPT2 correlation energy is computationally very expensive, although special numerical techniques have been developed to lower the computational cost.\cite{Sharma:2014p111101,Sokolov:2016p064102,Sharma:2016p034103,Sokolov:2017p244102}
Instead, in most calculations, the first-order wavefunction $\ket{\Psi^{(1)}_I}$ in \cref{eq:corr_energy} is approximated in the {\it contracted} form 
\begin{align}
	\label{eq:1st_order_wfn_contracted}
	\ket{\Psi^{(1)}_I} 
	\approx \sum_{\mu} t_{\mu I}^{(1)} \hat{O}_\mu \ket{\Psi_I^{(0)}}
	\equiv \sum_{\mu} t_{\mu I}^{(1)} \ket{\Phi_{\mu I}} \ ,
\end{align}
where $\ket{\Phi_{\mu I}}$ are many-particle basis functions called perturbers that are formed by acting the one- and two-electron excitation operators $ \hat{O}_\mu$ on the zeroth-order wavefunction $\ket{\Psi_I^{(0)}}$ (e.g., $ \hat{O}_\mu =\c{x} \a{i},\  \c{x} \c{y} \a{j} \a{i}, \ \c{a} \c{b} \a{x} \a{i}, \ \ldots$). 

Two contraction schemes have been developed, namely: (i) {\it strongly contracted} NEVPT2 (sc-NEVPT2) where only one perturber function is employed for each unique class of excitation operators $ \hat{O}_\mu$,\cite{Angeli:2001p10252,Angeli:2002p9138,Angeli:2006p054108} and (ii) {\it fully internally contracted} NEVPT2 (also known as {\it partially contracted} NEVPT2, pc-NEVPT2) where multiple perturbers are used for each excitation class.
While the strong contraction approximation simplifies the NEVPT2 implementation, it introduces non-negligible errors in the correlation energy\cite{Sokolov:2016p064102,Sivalingam:2016p054104,Sokolov:2017p244102} and suffers from the lack of orbital invariance with respect to the rotations within inactive orbital subspaces, which leads to the numerical instabilities in the evaluation of analytic gradients and properties.\cite{Guo:2016p094111,Park:2019p5417}
For this reason, in this work we will only consider the pc-NEVPT2 variant and will refer to it as NEVPT2 henceforth.

An attractive feature of NEVPT2 is the ability to avoid the intruder-state problems common in multireference theories\cite{Evangelisti:1987p4930,Angeli:2001p10252,Evangelista:2014p054109} by including the two-electron interaction term in the definition of zeroth-order Hamiltonian $\hat{\cal{H}}^{(0)}$ (\cref{eq:dyall_h_act}). 
Although the conventional (state-specific) NEVPT2 approach can be applied to ground and excited electronic states, it does not properly treat the interaction between states when they are very close to each other in energy, leading to the incorrect description of potential energy surfaces at conical intersections, avoided crossings, and in chemical systems with high density of states. 
A powerful approach to solve this problem is to employ the quasidegenerate formulation of NEVPT2 (QDNEVPT2), which is described in \cref{sec:theory:qdnevpt}.

\subsection{Quasidegenerate $N$-electron valence perturbation theory}
\label{sec:theory:qdnevpt}

In QDNEVPT2,\cite{Angeli:2004p4043} the energies of electronic states are computed by diagonalizing the matrix of effective Hamiltonian 
\begin{align}
	\label{eq:eff_H_eig_problem}
	\boldsymbol{\cal{H}}_{\mathbf{eff}}\, \textbf{Y} = \textbf{Y} \, \textbf{E}\ ,
\end{align}
which accounts for the coupling between model states $\ket{\Psi^{(0)}_I}$ after their perturbation (so-called ``diagonalize--perturb--diagonalize'' approach).\cite{Zaitsevskii:1995p597, Shavitt:2008p5711}
The original QDNEVPT2 method formulated by Angeli et al.\cite{Angeli:2004p4043} employs a non-Hermitian effective Hamiltonian matrix $\boldsymbol{\cal{H}}_{\mathbf{eff}}$ with elements
\begin{align}
	\label{eq:eff_H_nosym}
	\langle{\Psi_I^{(0)}}|\hat{\cal{H}}_{eff}|{\Psi_J^{(0)}}\rangle 
	&= E^{(0)}_I \delta_{IJ} 
	+ \langle{\Psi^{(0)}_{I}}|\hat{\cal{V}}|{\Psi^{(0)}_J}\rangle  \notag \\
	&+ \langle{\Psi^{(0)}_{I}}|\hat{\cal{V}}|{\Psi^{(1)}_J}\rangle  \ .
\end{align}
In \cref{eq:eff_H_nosym}, the first-order wavefunctions $\ket{\Psi^{(1)}_I}$ are approximated by \cref{eq:1st_order_wfn_contracted} where the contraction coefficients $t_{\mu I}^{(1)}$ are computed independently for each model state $\ket{\Psi^{(0)}_I}$ with energy $E^{(0)}_I$ obtained from a state-averaged CASSCF calculation (SA-CASSCF).\cite{Angeli:2001p10252,Angeli:2002p9138,Angeli:2004p4043,Zaitsevskii:1995p597}

An alternative formulation of QDNEVPT2 can be obtained from the Kirtman--Certain--Hirschfelder form of the canonical Van Vleck perturbation theory\cite{Kirtman:1981p798,Kirtman:2003p3890,Certain:2003p5977,Shavitt:2008p5711} where a Hermitian effective Hamiltonian is used:
\begin{align}
	\label{eq:eff_H_sym}
	\langle{\Psi_I^{(0)}}|\hat{\cal{H}}_{eff}|{\Psi_J^{(0)}}\rangle 
	&= E^{(0)}_I \delta_{IJ}  + \langle{\Psi^{(0)}_{I}}|\hat{\cal{V}}|{\Psi^{(0)}_J}\rangle  \notag \\
	&+ \frac{1}{2} \langle{\Psi^{(0)}_{I}}|\hat{\cal{V}}|{\Psi^{(1)}_J}\rangle  \notag \\
	&+ \frac{1}{2} \langle{\Psi^{(1)}_{I}}|\hat{\cal{V}}|{\Psi^{(0)}_J}\rangle  \ .
\end{align}
\cref{eq:eff_H_sym} was employed by Sharma et al.\@ in the implementation of uc-QDNEVPT2 with matrix product states\cite{Sharma:2016p034103} and can be seen as a symmetrized version of \cref{eq:eff_H_nosym}.
In practice, diagonalizing the effective Hamiltonians defined in \cref{eq:eff_H_nosym,eq:eff_H_sym} yields very similar electronic energies that differ by less than $10^{-5}$ $E_h$.
For this reason, in this work we will employ the symmetric formulation of QDNEVPT2, which simplifies the evaluation of excited-state properties and oscillator strengths.

For a fixed number of active orbitals, the computational cost of QDNEVPT2 scales as $\mathcal{O}(M^5)$ with the size of one-electron basis set ($M$). 
However, evaluating the matrix elements in \cref{eq:eff_H_sym} and the contraction coefficients $t_{\mu I}^{(1)}$ in \cref{eq:1st_order_wfn_contracted} requires computing the three-particle transition reduced matrices (3-TRDM, $\braket{\Psi^{(0)}_I|\c{u}\c{v}\c{w}\a{x}\a{y}\a{z}|\Psi^{(0)}_J}$, $I > J$) and the four-particle state-specific reduced density matrices (4-RDM, $\braket{\Psi^{(0)}_I|\c{u}\c{v}\c{w}\c{x}\a{x'}\a{w'}\a{v'}\a{u'}|\Psi^{(0)}_I}$)  in the active space with the computational cost scaling as $\mathcal{O}(N_{det}N_{states}^{2}N_{act}^6)$ and $\mathcal{O}(N_{det}N_{states}N_{act}^8)$, respectively, where $N_{det}$ is the number of Slater determinants in the complete active space, $N_{states}$ is the number of model states $\ket{\Psi^{(0)}_I}$, and $N_{act}$ is the number of active orbitals.

\subsection{Incorporating spin--orbit coupling in QDNEVPT2}
\label{sec:theory:so_qdnevpt}

To incorporate spin--orbit coupling into the QDNEVPT2 simulations of excited states, the effective nonrelativistic Hamiltonian in \cref{eq:eff_H_sym} must be augmented with the terms that describe the interaction between electronic spin and orbital angular momentum. 
These contributions can be derived by starting with the one-electron four-component Dirac Hamiltonian,\cite{Dyall:1995_book,Reiher:2014_book} incorporating two-electron interactions, and introducing approximations that transform the resulting Hamiltonian to a two-component form.\cite{Fleig:2012p2,Kutzelnigg:2012p16, Kutzelnigg:2005p241102,Cheng:2014p164107,Liu:2010p1679} 
Depending on how the transformation from four-component to two-component Hamiltonian is performed, different two-component spin--orbit Hamiltonians have been formulated.\cite{Saue:2011p3077,Reiher:2014_book,Kutzelnigg:2012p16, Kutzelnigg:2005p241102,Barysz:2002p2696,Dyall:1998p9618,Neese:1998p6568,Neese:2005p034107,Douglas:1974p89,Wolf:2002p9215,VanLenthe:1993p4597} 

In this work, we employ the Breit--Pauli (BP) Hamiltonian,\cite{Breit:1932p616,Mourad:1994p267,Fontana:2003p1357,Dyall:1995_book} which can be expressed as:
\begin{align}
	\label{eq:bp_h_sf_so}
	{\hat{\cal{H}}}_{BP} &= {\hat{\cal{H}}^{SF}}_{BP}  + {\hat{\cal{H}}^{SO}}_{BP} 
\end{align}
where ${\hat{\cal{H}}^{SF}}_{BP}$ and ${\hat{\cal{H}}^{SO}}_{BP}$ are the spin-free and spin--orbit contributions, respectively.
The ${\hat{\cal{H}}^{SF}}_{BP}$  term incorporates important scalar relativistic effects into the one-electron kinetic energy and electron nuclear attraction, which can be easily included by modifying the one-electron integrals in the CASSCF and QDNEVPT2 calculations. 
We will discuss the treatment of scalar relativistic effects in \cref{sec:implementation} and instead, here, will focus on the spin--orbit contribution to the BP Hamiltonian
\begin{align}
	\label{eq:bp_h}
	{\hat{\cal{H}}}^{SO}_{BP} 
	&= \sum_{\xi} \left( \sum_{i} \hat{h}_{\xi}(i) \cdot \hat{s}_{\xi}(i) \right. \notag \\
	&+\left. \sum_{i\neq j} [2\hat{g}_{\xi,soo}(i,j) + \hat{g}_{\xi,sso}(i,j)] \cdot \hat{s}_{\xi}(i) \right) \ , 
\end{align}
where $\hat{h}_{\xi}(i) \cdot \hat{s}_{\xi}(i)$ ($\xi = x, y, z$) is the one-electron spin--orbit operator of electron $i$
\begin{align}
	\label{eq:bp_h_1}
	\hat{h}_{\xi}(i) 
	&= \frac{1}{2c^{2}}\sum_{A}\frac{Z_{A}[\mathbf{r}_{iA}\times \mathbf{\hat{p}}(i)]_{\xi}}{r^{3}_{iA}} \ , 
\end{align}
while $\hat{g}_{\xi,soo}(i,j) \cdot \hat{s}_{\xi}(i)$ and $\hat{g}_{\xi,sso}(i,j) \cdot \hat{s}_{\xi}(i)$ are the so-called ``spin--other orbit'' and ``spin--same orbit'' two-electron terms, respectively:
\begin{align}
	\label{eq:bp_h_2}
	\hat{g}_{\xi,soo}(i,j) &= -\frac{1}{2c^{2}}\frac{[\mathbf{r}_{ij}\times \mathbf{\hat{p}}(j)]_{\xi}}{r^{3}_{ij}} \ , \\
	\label{eq:bp_h_3}
	\hat{g}_{\xi,sso}(i,j) &= -\frac{1}{2c^{2}} \frac{[\mathbf{r}_{ji}\times \mathbf{\hat{p}}(i)]_{\xi}}{r^{3}_{ij}} \ .
\end{align}
In \cref{eq:bp_h,eq:bp_h_1,eq:bp_h_2,eq:bp_h_3}, $Z_A$ denotes the nuclear charge on nucleus $A$, $\mathbf{r}_{ij}$ and $\mathbf{r}_{iA}$ are the relative coordinates of electron $i$ with respect to electron $j$ and nucleus $A$, respectively, $\mathbf{\hat{p}} (i)$ is the momentum operator of electron $i$, and $\hat{s}_{\xi}(i)$ is the $\xi$-component of the spin operator. 

The spin--orbit BP Hamiltonian in \cref{eq:bp_h} can be expressed in the second-quantized form:
\begin{align}
	 \label{eq:bp_h_second_quantized}
	{\hat{\cal{H}}}^{SO}_{BP} 
	&= \sum_{\xi}\left(\sum_{pq} h^{\xi}_{pq}\hat{D}^{\xi}_{pq} \right. \notag \\
	&+ \left. \sum_{pqrs} [2g^{\xi,soo}_{pqrs} + g^{\xi,sso}_{pqrs}]\hat{D}^{\xi}_{pqrs}\right) \ ,
\end{align}
where $\hat{D}^{\xi}_{pq}$ and $\hat{D}^{\xi}_{pqrs}$ are the one- and two-electron spin excitation operators, 
\begin{align}
	\hat{D}^{x}_{pq} &= \frac{1}{2}({a}^{\dagger}_{p\alpha}{a}_{q\beta} + {a}^{\dagger}_{p\beta}{a}_{q\alpha}) \ , \\
	\hat{D}^{y}_{pq} &= \frac{i}{2}({a}^{\dagger}_{p\beta}{a}_{q\alpha} - {a}^{\dagger}_{p\alpha}{a}_{q\beta}) \ , \\
	\hat{D}^{z}_{pq} &= \frac{1}{2}({a}^{\dagger}_{p\alpha}{a}_{q\alpha} - {a}^{\dagger}_{p\beta}{a}_{q\beta}) \ , \\
	\hat{D}^{\xi}_{pqrs} &= {a}^{\dagger}_{r\alpha}\hat{D}^{\xi}_{pq}{a}_{s\alpha} + {a}^{\dagger}_{r\beta}\hat{D}^{\xi}_{pq}{a}_{s\beta} \ ,
\end{align}
while $h^{\xi}_{pq}$, $g^{\xi,soo}_{pqrs}$, and $g^{\xi,sso}_{pqrs}$  are the one- and two-electron integrals calculated in the spatial molecular orbital basis ($\phi_p$):
\begin{align}
	 h^{\xi}_{pq} &= \langle{\phi_p(1)}| \hat{h}_{\xi}(1) |{\phi_q(1)}\rangle \ , \\
	 \label{eq:g_soo}
	g^{\xi,soo}_{pqrs} &= \langle{\phi_p(1) \phi_r (2)}| \hat{g}_{\xi,soo}(1,2) |{\phi_q (1)\phi_s (2)}\rangle \ , \\
	 \label{eq:g_sso}
	g^{\xi,sso}_{pqrs} &= \langle{\phi_p(1) \phi_r (2)}| \hat{g}_{\xi,sso}(1,2) |{\phi_q (1)\phi_s (2)}\rangle \ .
\end{align}
The spin--other orbit and spin--same orbit two-electron integrals in \cref{eq:g_soo,eq:g_sso} are related to each other via a permutation: $g^{\xi,soo}_{pqrs} = g^{\xi,sso}_{rspq} \equiv g^{\xi}_{pqrs}$.
Thus, using the Hamiltonian in \cref{eq:bp_h_second_quantized} requires calculating only one set of these spin--orbit two-electron integrals.

Treating ${\hat{\cal{H}}}^{SO}_{BP}$ as a perturbation to the nonrelativistic Hamiltonian $\hat{\cal{H}}$ (\cref{eq:hamiltonian}), we modify the QDNEVPT2 effective Hamiltonian as follows
\begin{align}
	\label{eq:eff_H_sym_SO}
	\langle{\Psi_I^{(0)}}|\hat{\cal{H}}_{eff}^{SO}|{\Psi_J^{(0)}}\rangle 
	&= E^{(0)}_I \delta_{IJ} \notag \\
	&+ \langle{\Psi^{(0)}_{I}}|\hat{\cal{V}} + {\hat{\cal{H}}}^{SO}_{BP}|{\Psi^{(0)}_J}\rangle  \notag \\
	&+ \frac{1}{2} \langle{\Psi^{(0)}_{I}}|\hat{\cal{V}}|{\Psi^{(1)}_J}\rangle  \notag \\
	&+  \frac{1}{2} \langle{\Psi^{(1)}_{I}}|\hat{\cal{V}}|{\Psi^{(0)}_J}\rangle \ .
\end{align}
Diagonalizing $\boldsymbol{\cal{H}}_{\mathbf{eff}}^{\mathbf{SO}}$ in \cref{eq:eff_H_sym_SO} incorporates the spin--orbit coupling effects up to the first order in perturbation theory and will be referred to as the SO-QDNEVPT2 approach.
 
\subsection{Spin--orbit mean-field approximation in SO-QDNEVPT2}

Including the spin--orbit term in \cref{eq:eff_H_sym_SO} does not increase the computational scaling of QDNEVPT2 with the system size, but requires an expensive calculation and storage of all spin--orbit two-electron integrals, $g^{\xi}_{pqrs}$.
Since the one- and two-electron terms in the BP Hamiltonian (\cref{eq:bp_h}) have opposite signs, neglecting the $g^{\xi}_{pqrs}$ contributions can lead to a significant overestimation of spin--orbit coupling energies.
Alternatively, incorporating the spin--orbit coupling effects can be simplified by invoking the spin--orbit mean-field approximation (SOMF),\cite{Hess:1996p365,Berning:2000p1823} which describes the two-electron spin--orbit interactions in a way analogous to the mean-field treatment of electronic repulsion in Hartree--Fock theory.
The SOMF approximation has been used to incorporate spin--orbit coupling in a variety of electronic structure theories with a wide range of applications.\cite{Hess:1996p365,Malmqvist:2002p230,Ganyushin:2013p104113,Epifanovsky:2015p64102,Mussard:2018p154,Meitei:2020p3597,Netz:2021p5530} 

Within the SOMF approximation, the spin--orbit BP Hamiltonian (\cref{eq:bp_h_second_quantized}) can be expressed as an effective one-electron operator
\begin{align}
	\label{eq:bp_h_second_quantized_somf}
	{\hat{\cal{H}}}^{SOMF}_{BP} 
	&= \sum_{\xi} \sum_{pq} F^{\xi}_{pq}\hat{D}^{\xi}_{pq} 
\end{align}
with matrix elements 
\begin{align}
F^{\xi}_{pq} = h^{\xi}_{pq} + \sum_{rs} \Gamma_{r}^{s} \left(g^{\xi}_{rspq} - \frac{3}{2}g^{\xi}_{prsq} + \frac{3}{2}g^{\xi}_{qrsp}\right) \ ,
\end{align}
where $\Gamma_{r}^{s} = \gamma_{r\alpha}^{s\alpha} + \gamma_{r\beta}^{s\beta}$ is the spinless one-particle reduced density matrix calculated with respect to the SA-CASSCF wavefunction. 
Replacing ${\hat{\cal{H}}}^{SO}_{BP}$ in \cref{eq:eff_H_sym_SO} with ${\hat{\cal{H}}}^{SOMF}_{BP}$ defines the SOMF-approximated QDNEVPT2 effective Hamiltonian
\begin{align}
	\label{eq:eff_H_sym_SOMF}
	\langle{\Psi_I^{(0)}}|\hat{\cal{H}}_{eff}^{SOMF}|{\Psi_J^{(0)}}\rangle 
	&= E^{(0)}_I \delta_{IJ} \notag \\
	&+ \langle{\Psi^{(0)}_{I}}|\hat{\cal{V}} + {\hat{\cal{H}}}^{SOMF}_{BP}|{\Psi^{(0)}_J}\rangle  \notag \\
	&+ \frac{1}{2} \langle{\Psi^{(0)}_{I}}|\hat{\cal{V}}|{\Psi^{(1)}_J}\rangle \notag \\
	&+ \frac{1}{2} \langle{\Psi^{(1)}_{I}}|\hat{\cal{V}}|{\Psi^{(0)}_J}\rangle \ ,
\end{align}
which we will abbreviate as SOMF-QDNEVPT2.

\section{Implementation}
\label{sec:implementation}

We implemented the SO-QDNEVPT2 and SOMF-QDNEVPT2 methods in \textsc{Prism}, which is a Python program for excited-state and spectroscopic simulations that is being developed in our group. 
The \textsc{Prism} code is interfaced with the \textsc{Pyscf} software package\cite{Sun:2020p024109} to obtain the one- and two-electron integrals, as well as the SA-CASSCF molecular orbitals and model state wavefunctions.
Here, we provide additional details regarding the SO-QDNEVPT2 and SOMF-QDNEVPT2 implementations developed in this work.

\textit{1. Treating scalar relativistic effects.}
As discussed in \cref{sec:theory:so_qdnevpt}, describing spin--orbit coupling must be accompanied with a treatment of spin-free (scalar) relativistic effects, which can be incorporated variationally by modifying the one-electron integrals in the SA-CASSCF and QDNEVPT2 calculations. 
Although the scalar relativistic effects can be treated using the spin-free part of the BP Hamiltonian (${\hat{\cal{H}}^{SF}}_{BP}$ in \cref{eq:bp_h_sf_so}), in our implementation of SO-QDNEVPT2 and SOMF-QDNEVPT2 we employ the spin-free exact two-component (X2C) Hamiltonian,\cite{Kutzelnigg:2005p241102,Liu:2006p044102,Peng:2007p104106,Ilias:2007p064102,Liu:2009p031104} which offers a more rigorous treatment of scalar relativistic effects with a minor increase in computational cost. 
This approach has been successfully used in other implementations utilizing approximate two-component spin--orbit Hamiltonians.\cite{Li:2014p054111, Liu:2018p144108, Mussard:2018p154} 

\textit{2. Avoiding the calculation of 4-RDM.}
As mentioned in \cref{sec:theory:qdnevpt}, evaluating the contraction coefficients $t_{\mu I}^{(1)}$ of the first-order QDNEVPT2 wavefunctions (\cref{eq:1st_order_wfn_contracted}) requires to calculate and store 4-RDM, which is prohibitively expensive for large active spaces. 
In our implementation of SO-QDNEVPT2 and SOMF-QDNEVPT2, we avoid computing 4-RDM without introducing any approximations using the approach developed in Ref.\@ \citenum{Chatterjee:2020p6343}.
This allows to greatly reduce disk and memory storage while lowering the computational scaling of our implementation to $\mathcal{O}(N_{det}N_{states}^{2}N_{act}^6)$ with the number of active orbitals $N_{act}$.

\textit{3. Preserving the degeneracy of internally contracted states.}
The internal contraction approximation employed in QDNEVPT2 can result in small errors violating the degeneracy of spin--orbit-coupled states in open-shell systems with high symmetry (e.g., isolated atoms, linear molecules, etc.). 
These errors originate from using the multipartitioning technique\cite{Zaitsevskii:1995p597} in QDNEVPT2 where the contraction coefficients $t_{\mu I}^{(1)}$ in \cref{eq:1st_order_wfn_contracted} are determined independently for each model state $\ket{\Psi^{(0)}_I}$.
If two or more SA-CASSCF model states $\ket{\Psi^{(0)}_I}$ have the same energies, small differences in internal contraction for each of these states can result in lifting of their degeneracy at the QDNEVPT2 level of theory.
These errors also emerge in the SO-QDNEVPT2 calculations breaking the degeneracy of spin--orbit-coupled states.
To prevent this, for each set of $N_{deg}$ degenerate SA-CASSCF model states $\ket{\Psi^{(0)}_I}$ we compute $t_{\mu I}^{(1)}$ with respect to a state-averaged model wavefunction
\begin{align}
	\label{eq:mod_state_averaging}
	\ket{\tilde{\Psi}^{(0)}} = \frac{1}{N_{deg}} \sum_{I}^{deg} \ket{\Psi^{(0)}_I} \ ,
\end{align}
where the summation is restricted to model states $\ket{\Psi^{(0)}_I}$ with the same energy $E^{(0)}_I$.
Note that state-averaging in \cref{eq:mod_state_averaging} is used only for evaluating $t_{\mu I}^{(1)}$ (i.e., describing dynamical correlation) and not for computing the matrix elements of effective Hamiltonian.  
As demonstrated in Supplementary Information, using this approach allows to fully restore the degeneracy of spin--orbit-coupled states while taking advantage of internal contraction without affecting the accuracy of SO-QDNEVPT2 and SOMF-QDNEVPT2.

\textit{4. Calculating oscillator strengths.}
Our implementation of SO-QDNEVPT2 and SOMF-QDNEVPT2 is also capable of computing oscillator strengths according to the following equation:
\begin{align}
	\label{eq:osc_strength}
	f^{osc}_{if} = \frac{2}{3} (E_f- E_i) \sum_{ \xi pq IJ} \left| \mu_{pq}^{\xi} Y^*_{If} \Gamma_{pq}^{IJ} Y_{Ji}\right|^2 \ ,
\end{align}
where $\Gamma_{pq}^{IJ}$ is the spinless 1-TRDM computed with respect to the model states $\ket{\Psi^{(0)}_I}$ and $\ket{\Psi^{(0)}_J}$, $\mu_{pq}^{\xi}$ are the dipole moment integrals calculated in the spatial molecular orbital basis, while $E_k$ and $Y_{Jk}$ are the eigenvalues and eigenvectors of SO-QDNEVPT2 or SOMF-QDNEVPT2 effective Hamiltonian for the initial ($k = i$) and final ($k = f$) electronic states.

\section{Computational details}
\label{sec:comp_details}

\begin{table*}[t!]
	\caption{Spin--orbit zero-field splitting (\cm) in the ${}^2\Pi$ ground states of GeH and SnH computed using SO-QDNEVPT2 and SOMF-QDNEVPT2 with the (5e, 5o) active space averaging over both spatial components of  ${}^2\Pi$ in SA-CASSCF. Results are compared to the variational two-component calculations using X2C-MRPT2\cite{Lu:2022p2983} and available experimental data.\cite{huber:2013p298}
	Oscillator strengths computed using SO-QDNEVPT2 and SOMF-QDNEVPT2 are given in parentheses.
	}
	\label{tab:gr14}
	\setstretch{1}
	\small
	\centering
	\begin{threeparttable}
		\begin{tabular}{c c c c c} 
			\hline\hline
			Molecule & SO-QDNEVPT2 & SOMF-QDNEVPT2 & X2C-MRPT2\cite{Lu:2022p2983} & Experiment\cite{huber:2013p298} \\ 
			\hline
			GeH& 869.9 		& 870.0 		&  898.6	& 892.5\\ 
			& (0.0119)		& (0.0119)		&  	& \\ 
			SnH& 2372.9 	& 2373.0 		& 2197.5	& 2178.9  \\
			&  (0.0435) 	& (0.0435)		& &  \\
			\hline\hline
		\end{tabular}
	\end{threeparttable}
\end{table*}

We benchmarked the SO-QDNEVPT2 and SOMF-QDNEVPT2 methods for a variety of atoms and small molecules, namely: i) group 14 hydrides (\ce{GeH} and \ce{SnH}, \cref{sec:results:group_14}); ii) group 16 hydrides (from \ce{OH} to \ce{TeH}, \cref{sec:results:group_16}); iii) $3d$ and $4d$ transition metal ions with the 2+ charge (\cref{sec:results:3d_4d}); and iv) actinyl oxide ions (\ce{NpO2^2+} and \ce{PuO2^2+}, \cref{sec:results:Np_Pu}).

In \cref{sec:results:group_14}, we study the spin--orbit splitting in the ${}^2\Pi$ ground electronic states of \ce{GeH} and \ce{SnH} and its dependence on the parameters of SA-CASSCF calculations, such as the active space size, number of CASCI states, and weights used for state-averaging. 
All calculations of \ce{GeH} and \ce{SnH} were performed using the all-electron X2C-TZVPall-2c basis set\cite{Pollak:2017p3696}. 
We considered two different active spaces: 5 electrons in 5 active orbitals (5e, 5o) and 15 electrons in 10 active orbitals (15e, 10o).
The (5e, 5o) active space included two $\sigma$, two $\pi$, and one $\sigma^*$ orbitals.
The (15e, 10o) active space incorporated additional five $d$ orbitals ($3d$ for \ce{GeH} or $4d$ for \ce{SnH}). 
Since ${}^2\Pi$ is spatially doubly degenerate, the SA-CASSCF calculations were performed by averaging over the two lowest-energy states.
Experimental bond lengths of 1.5880 $\angstrom$ for GeH and 1.7815 $\angstrom$ for SnH were used in all calculations.\cite{Linstrom:2001p1059} 

For the group 16 hydrides (\cref{sec:results:group_16}), we investigate the dependence of ${}^2\Pi$ ground-state spin--orbit splitting on the basis set. 
In this study, we use the Dunning's correlation consistent basis sets\cite{Dunning:1989p1007,Wilson:1996p339,Woon:1998p1358,Wilson:1999p7667} cc-pV$X$Z ($X$ = T, Q, 5), fully uncontracted cc-pV$X$Z (unc-cc-pV$X$Z), and the ANO-RCC basis developed by Roos et al.\cite{Roos:2005p6575, Roos:2009p87} 
For \ce{TeH}, the DK3 variants of the cc-pV$X$Z basis sets were used for the Te atom (cc-pV$X$Z-DK3, $X$ = T, Q).\cite{Hill:2017p244106}
The active space was comprised of two $\sigma$, two $\pi$, and one $\sigma^*$ molecular orbitals (7e, 5o).
As for the group 14 hydrides, two CASCI states were averaged in SA-CASSCF.
All computations were carried out using the experimental bond lengths:\cite{huber:2013p298, Ram:2000p9, Fink:1989p19} $r$\textsubscript{\ce{OH}} = 0.96966 $\angstrom$, $r$\textsubscript{\ce{SH}} = 1.3409 $\angstrom$ , $r$\textsubscript{\ce{SeH}} = 1.4643 $\angstrom$, and $r$\textsubscript{\ce{TeH}} = 1.65587 $\angstrom$.

In \cref{sec:results:3d_4d}, we use our implementation of SO-QDNEVPT2 and SOMF-QDNEVPT2 to study the spin--orbit coupling in the ground and excited electronic states of $3d$ and $4d$ transition metal ions with the 2+ charge.
The active spaces of $3d$ metal ions included: $3d$ and $4d$ orbitals for \ce{V^2+}, \ce{Cr^2+}, and \ce{Co^2+}; $3d$, $4d$, and $4s$ orbitals for \ce{Ti^2+}, \ce{Fe^2+}, \ce{Ni^2+} and \ce{Cu^2+}; and $3d$, $4d$, $4s$, and $4p$ orbitals for \ce{Sc^2+}. 
For the $4d$ metal ions, we used the same active spaces as for the $3d$ ions within each group of periodic table, but with the principal quantum number of each active orbital increased by one. 
All calculations of $3d$ and $4d$ metal ions used the Sapporo-TZP\cite{Noro:2012p1} basis set.
The SA-CASSCF calculations were performed by averaging over several electronic states, as described in the Supplementary Information.

Finally, in \cref{sec:results:Np_Pu}, we present the results of SO-QDNEVPT2 and SOMF-QDNEVPT2 calculations for linear \ce{NpO2^2+} and \ce{PuO2^2+} using the ANO-RCC-VTZP basis. 
The structural parameters were obtained from Refs.\@ \citenum{Gendron:2014p7994} and \citenum{Gendron:2014p8577}: $r$\textsubscript{NpO} = 1.70 $\angstrom$ and $r$\textsubscript{PuO} = 1.682 $\angstrom$.
We employed the (7e, 10o) active space for \ce{NpO2^2+} and (8e, 10o) active space for \ce{PuO2^2+} (see Supplementary Information for details). 
The SA-CASSCF calculations were performed by averaging over 25 and 26 CASCI states for \ce{NpO2^2+} and \ce{PuO2^2+}, respectively.

\section{Results}
\label{sec:results}

\subsection{Spin--orbit coupling in group 14 hydrides and its dependence on the parameters of SA-CASSCF calculations}
\label{sec:results:group_14}

\begin{figure*}[t!]
 	\subfigure[]{
 		\includegraphics[width=0.48\textwidth]{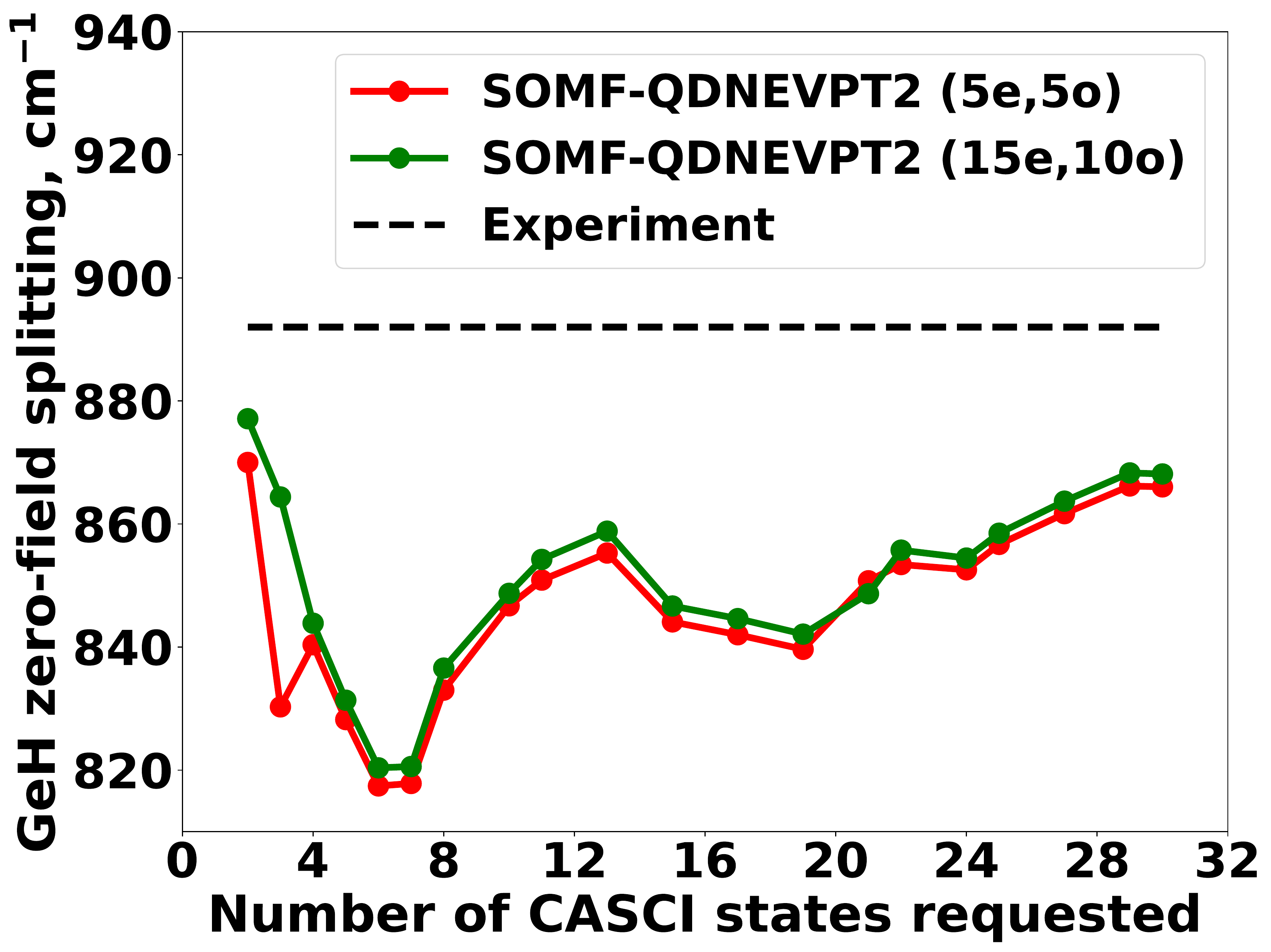}\label{fig:gr14_a}}  
 	\subfigure[]{
 		\includegraphics[width=0.48\textwidth]{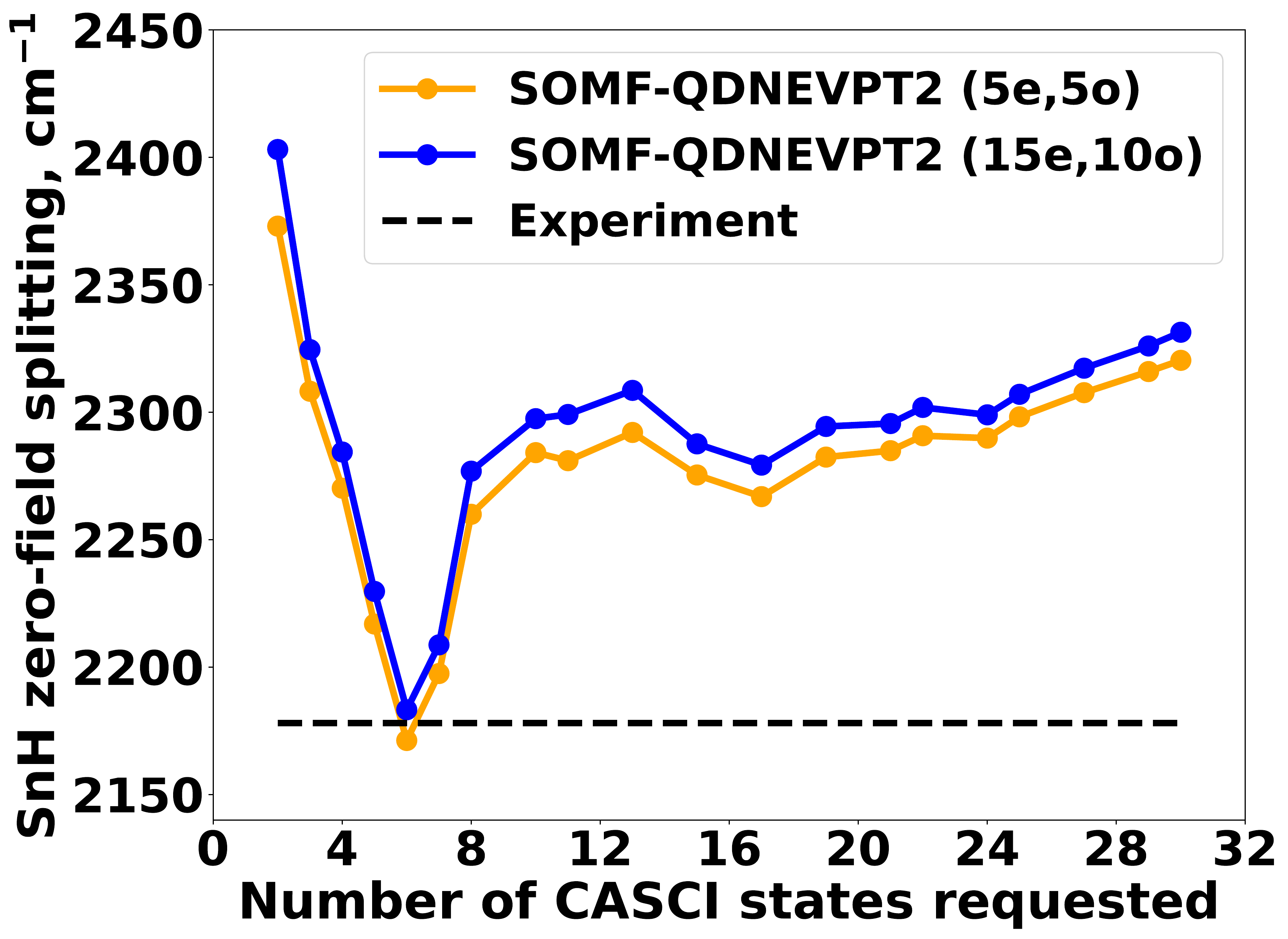}\label{fig:gr14_b}} 
	\subfigure[]{
		\includegraphics[width=0.48\textwidth]{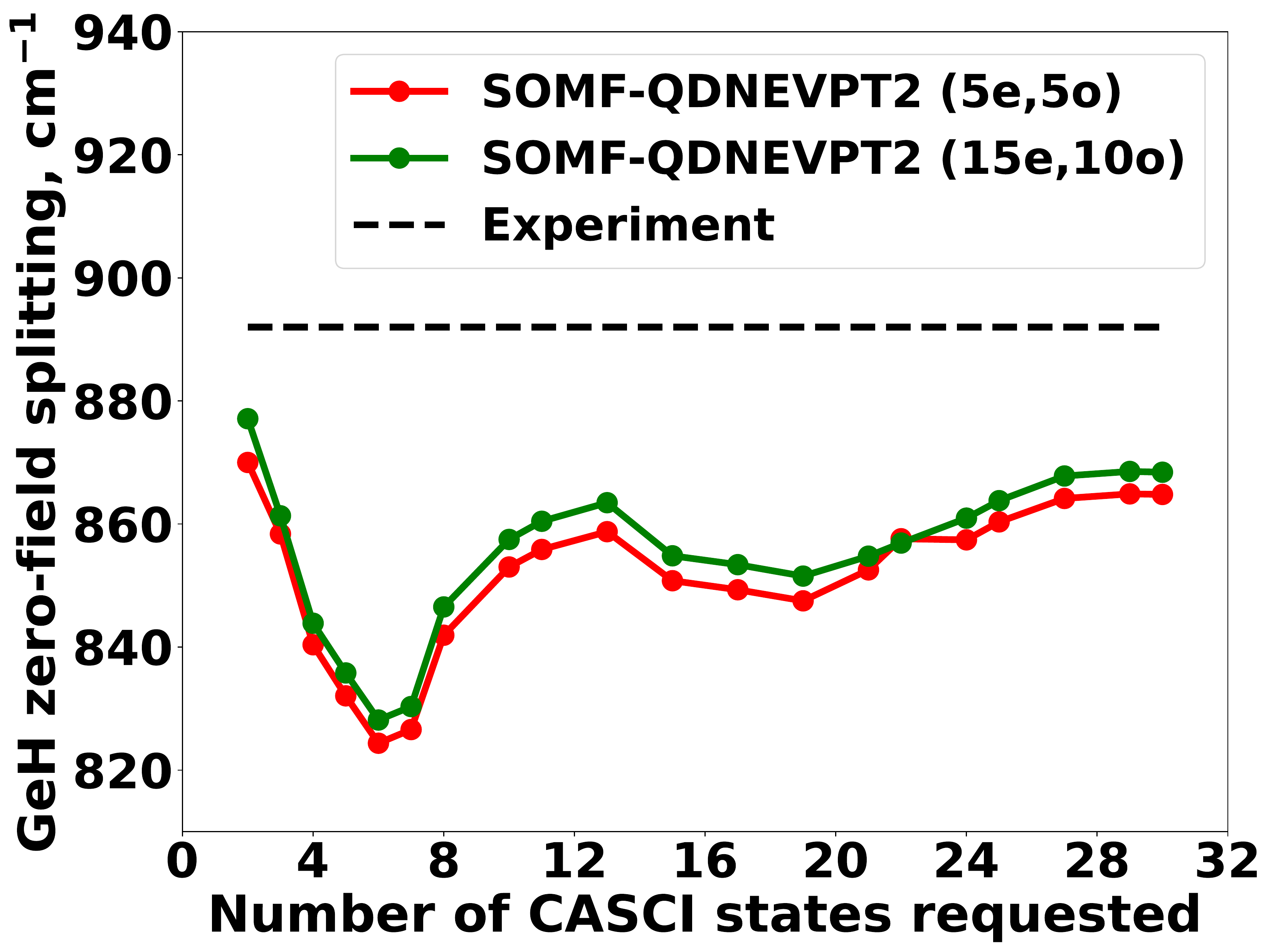}\label{fig:gr14_c}}  
	\subfigure[]{
		\includegraphics[width=0.48\textwidth]{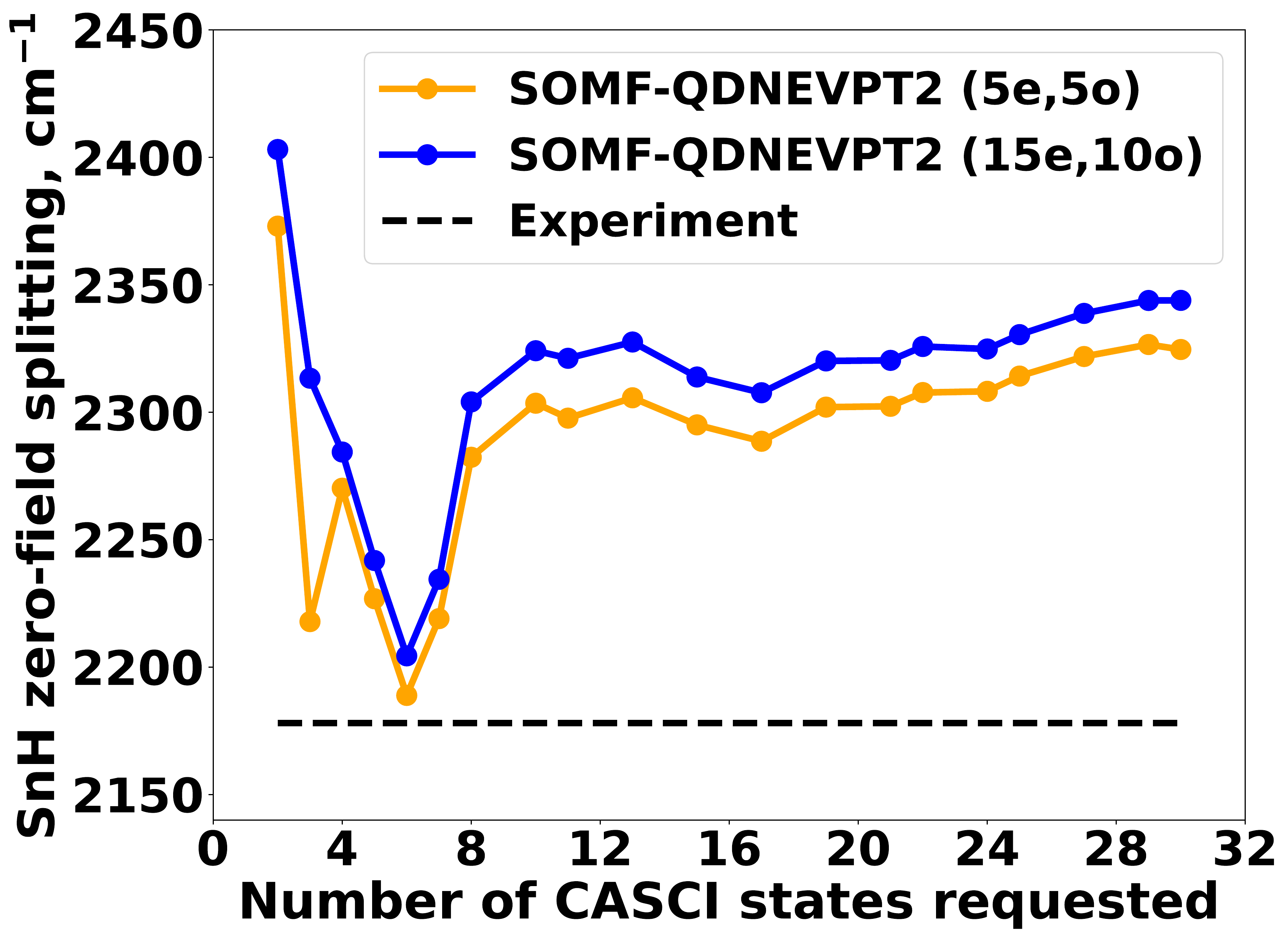}\label{fig:gr14_d}}  
 	\captionsetup{justification=raggedright,singlelinecheck=false}
 	\caption{Spin--orbit zero-field splitting in the ${}^2\Pi$ ground states of GeH (plots a and c) and SnH (plots b and d) computed using SOMF-QDNEVPT2 as the number of CASCI states included in SA-CASSCF and QDNEVPT2 increases. Results are shown for two active spaces: (5e, 5o) and (15e, 10o). In plots a and b, all CASCI states were assigned identical weights in state-averaging. In plots c and d, the weight of ${}^2\Pi$ ground state was fixed at 50\%, while the other states were assigned identical weights. 
    }
 	\label{fig:gr14}
\end{figure*}

We begin by investigating the accuracy of SO-QDNEVPT2 and SOMF-QDNEVPT2 for predicting the energy of spin--orbit zero-field splitting (ZFS) in the ${}^2\Pi$ ground states of GeH and SnH.
\cref{tab:gr14} shows the ZFS calculated using the (5e, 5o) active space with the two spatial components of ${}^2\Pi$ state averaged in SA-CASSCF for each molecule. 
The results of SO-QDNEVPT2 and SOMF-QDNEVPT2 with the first-order BP perturbative treatment of spin--orbit coupling are compared to the data from variational two-component X2C-MRPT2 calculations performed using the same basis set and molecular geometries by Lu et al.\cite{Lu:2022p2983}.
\cref{tab:gr14} also includes the SO-QDNEVPT2 and SOMF-QDNEVPT2 oscillator strengths and the available experimental data for comparison.\cite{huber:2013p298} 

For both molecules, the ZFS computed using SO-QDNEVPT2 and SOMF-QDNEVPT2 differ by only 0.1 \cm, suggesting that the SOMF approximation is very accurate in these systems. 
For GeH, the QDNEVPT2 methods are in a close agreement with the experiment underestimating ZFS by $\sim$ 22 \cm (2.5 \% error).
Larger errors (8.9 \%) are observed for SnH where the QDNEVPT2 methods overestimate ZFS by $\sim$ 194 \cm.
As expected, the oscillator strength of ${}^2\Pi_{1/2} \rightarrow {}^2\Pi_{3/2}$ transition increases with the increasing magnitude of spin--orbit coupling from GeH to SnH.
The X2C-MRPT2 method shows the smallest errors relative to experiment ($<$ 20 \cm, 0.8 \%), suggesting that the variational X2C treatment of spin--orbit coupling is important for very accurate predictions of ZFS in SnH.\cite{Cheng:2018p044108} 

We now analyze how the ground-state ZFS of GeH and SnH computed using SOMF-QDNEVPT2 depend on the parameters of SA-CASSCF calculations, namely: 1) the size of active space, 2) the number of CASCI states included in SA-CASSCF and QDNEVPT2 model space, and 3) the weights used in state-averaging.
\cref{fig:gr14_a,fig:gr14_b} show the variation in ${}^2\Pi$ ZFS of GeH (a) and SnH (b) calculated by increasing the number of CASCI states ($N_{states}$) from 2 to 30 with identical state-averaging weights for two active spaces: (5e, 5o) and (15e, 10o).
Similar trends are observed for both molecules.
As $N_{states}$ increases from 2 to 6, the computed ZFS decreases sharply by 7 to 10 \%.
Upon addition of four more CASCI states ($N_{states}$ = 10), ZFS increases by $\sim$ 3 to 5 \%.
Further increasing $N_{states}$ from 10 to 30 results in a slow increase of ZFS to a value that is just 2 to 3 \% lower than the ZFS for $N_{states}$ = 2.
However, up to $N_{states}$ = 30, the dependence of ZFS on the number of CASCI states does not level off.
In contrast to strong dependence on $N_{states}$, the computed ZFS does not change significantly with increasing active space in most calculations, except for GeH with $N_{states}$ = 3. 

To assess the dependence of ZFS on state-averaging weights, we performed the SOMF-QDNEVPT2 calculations by assigning the ${}^2\Pi$ ground state a weight of 50\% and distributing the other 50 \% weight equally among the remaining CASCI states. 
The ZFS calculated using this approach are shown in \cref{fig:gr14_c,fig:gr14_d} for GeH and SnH, respectively.
Except for $N_{states}$ = 3, the results of these calculations are very close to the SOMF-QDNEVPT2 calculations with equal weights for all CASCI states (\cref{fig:gr14_a,fig:gr14_b}).  

Overall, our results suggest that the ZFS calculated using SO-QDNEVPT2 and SOMF-QDNEVPT2 are more sensitive to the number of CASCI states included in SA-CASSCF and QDNEVPT2 than the state-averaging weights assigned to the individual states. 
While the calculations of ZFS in GeH and SnH have shown weak active-space dependence, we expect that the size of active space may be an important parameter for other systems where the electron correlation effects are more significant. 

\subsection{Spin--orbit coupling in group 16 hydrides and its basis set dependence}
\label{sec:results:group_16}

We now turn our attention to group 16 hydrides (\ce{OH}, \ce{SH}, \ce{SeH}, and \ce{TeH}), which are commonly used for the benchmark of electronic structure theories incorporating relativistic effects.\cite{Berning:2000p1823,Epifanovsky:2015p64102,Cheng:2018p044108,Meitei:2020p3597}
In this section, our focus is to investigate the dependence of ZFS in the ground ${}^2\Pi$ state of these systems on the choice of one-electron basis set. 
Our study employs three Dunning's correlation consistent basis sets\cite{Wilson:1996p339,Peterson:2002p10548} cc-pV$X$Z ($X$ = T, Q, 5), fully uncontracted cc-pV$X$Z (unc-cc-pV$X$Z), and the ANO-RCC basis developed by Roos et al.\cite{Roos:2005p6575, Roos:2009p87} 
For the Te atom in \ce{TeH}, we use the DK3 variants of cc-pV$X$Z basis sets (cc-pV$X$Z-DK3, $X$ = T, Q).\cite{Hill:2017p244106}

\begin{table*}[t!]
	\caption{
		Spin--orbit zero-field splitting (\cm) in the ${}^2\Pi$ ground states of group 16 hydrides computed using SO-QDNEVPT2 and SOMF-QDNEVPT2 with the (7e, 5o) active space averaging over both spatial components of  ${}^2\Pi$ in SA-CASSCF. 
		Results are compared to the calculations using RAS(SD)-1SF method\cite{Meitei:2020p3597} and available experimental data.\cite{huber:2013p298, Ram:2000p9, Fink:1989p19} 
		Oscillator strengths computed using SO-QDNEVPT2 and SOMF-QDNEVPT2 are given in parentheses.
	}
	\label{tab:gr16}
	\setstretch{1}
	\footnotesize
	\centering
	\begin{threeparttable}
		\begin{tabular}{cccccc}
			\hline\hline
			Molecule & Basis set   & SO-QDNEVPT2              & SOMF-QDNEVPT2            & RAS(SD)-1SF\cite{Meitei:2020p3597} & Experiment\cite{huber:2013p298, Ram:2000p9, Fink:1989p19} \\ \hline
			OH    & cc-pVTZ     	 & 137.0 (0.0003)      	& 135.8 (0.0003)                   & 134.4                         &                                            \\
			& cc-pVQZ       & 139.3 (0.0003)    		& 138.2 (0.0003)                & 137.4                         &                                               \\
			& cc-pV5Z       & 140.9 (0.0003)    		& 139.7 (0.0003)                             & 139.0                         &                                               \\
		    & unc-cc-pVTZ     	 & 137.3 (0.0003)      	& 136.2 (0.0003)                   & 135.3                        &                                            \\
			& unc-cc-pVQZ     	 & 139.6 (0.0003)      	& 138.5 (0.0003)                   & 137.6                        &                                            \\
            & unc-cc-pV5Z     	 & 141.1 (0.0003)      	& 139.9 (0.0003)                   & 139.1                         &                                            \\
			& ANO-RCC       & 141.1 (0.0003)    		& 139.9 (0.0003)                             & 134.9                         &   139                                            \\ 
			SH    & cc-pVTZ     	 & 350.0 (0.0025)      	& 349.8 (0.0025)                               & 360.7                         &                                            \\
			& cc-pVQZ       & 349.3 (0.0025)    		& 349.0 (0.0025)                              & 362.1                         &                                               \\
            & cc-pV5Z       & 354.7 (0.0026)    		& 354.5 (0.0026)                              & 392.7                         &                                               \\
            & unc-cc-pVTZ     	 & 355.3 (0.0026)      	& 355.1 (0.0026)                               & 384.0                         &                                            \\
			& unc-cc-pVQZ     	 & 356.3 (0.0026)      	& 356.1 (0.0026)                               & 387.8                         &                                            \\
            & unc-cc-pV5Z     	 & 356.4 (0.0026)      	& 356.2 (0.0026)                               & 390.0                         &                                            \\
			& ANO-RCC       & 356.0 (0.0027)    		& 355.8 (0.0027)                             & 354.3                         &  377                                             \\ 
			SeH   & cc-pVTZ     	 & 1544.1 (0.0149)   	& 1544.0 (0.0149)                             & 1603.0                        &                                           \\
			& cc-pVQZ        & 1542.5 (0.0151)  		& 1542.4 (0.0151)                            & 1634.1                        &                                               \\
			& cc-pV5Z       & 1585.1 (0.0155)   		& 1584.9 (0.0155)                             & 1711.6                        &                                               \\
            & unc-cc-pVTZ     	 & 1761.5 (0.0172)   	& 1761.4 (0.0171)                             & 1718.5                        &                                           \\
			& unc-cc-pVQZ     	 & 1765.9 (0.0173)   	& 1765.8 (0.0173)                             & 1729.6                        &                                           \\   
            & unc-cc-pV5Z     	 & 1766.9 (0.0174)  	& 1766.7 (0.0174)                            &                     &                                           \\
			& ANO-RCC       & 1773.1 (0.0175)  		& 1773.0 (0.0175)                             & 1828.2                        &  1763                                             \\ 
			TeH   & cc-pVTZ-DK3	 & 4294.6 (0.0593)      & 4294.5 (0.0593)                            &                              &                                           \\
			& cc-pVQZ-DK3    & 4290.3 (0.0595)  		& 4290.2 (0.0595)                             &                              &                                               \\
			& ANO-RCC       & 4284.1 (0.0596)   		& 4284.0 (0.0596)                             & 4602.3                        &     3816                                          \\ \hline\hline
		\end{tabular}
	\end{threeparttable}
\end{table*}

\cref{tab:gr16} compares the ${}^2\Pi$ ZFS and oscillator strengths computed using SO-QDNEVPT2 and SOMF-QDNEVPT2 with the data from the RAS(SD)-1SF method\cite{Meitei:2020p3597} and experiments.\cite{huber:2013p298, Ram:2000p9, Fink:1989p19}
For each molecule and basis set, the results of SO-QDNEVPT2 and SOMF-QDNEVPT2 are within 2 \cm of each other, demonstrating the high accuracy of SOMF approximation. 
For OH and SH, the simulated ZFS and oscillator strengths show weak basis set dependence.
In this case, the ZFS calculated using the five-zeta correlation consistent basis sets (cc-pV5Z and unc-cc-pV5Z) and the ANO-RCC basis set optimized for the calculations with relativistic Hamiltonians agree within 2 \cm of each other and deviate by less than 21 \cm from the experiment.

A different situation is observed for SeH where the changes in ZFS and oscillator strengths accelerate with the increasing cardinal number $X$ in cc-pV$X$Z, suggesting that the results computed using the contracted correlation consistent basis sets that are not optimized for calculations incorporating relativistic effects are far from the basis set limit.
This is further supported by the results computed using unc-cc-pV$X$Z, which show significantly larger ZFS (by $\sim$ 200 \cm) and much weaker dependence on the cardinal number $X$.
The ZFS computed using unc-cc-pV5Z (1767 \cm) is in a close agreement with the ZFS from ANO-RCC (1773 \cm) and experiment (1763 \cm).
Similar basis set dependence of ZFS is observed in the RAS(SD)-1SF data calculated by Meitei et al.\cite{Meitei:2020p3597}
For TeH, using the cc-pV$X$Z-DK3 basis sets ($X$ = T and Q) recontracted for relativistic calculations yields the ZFS values (4295 and 4290 \cm) that are similar to the ZFS computed with ANO-RCC (4284 \cm), which overestimates the experimental spin--orbit splitting by 468 \cm (12.2 \% error).

For all group 16 molecules, the ZFS computed using SO-QDNEVPT2 and SOMF-QDNEVPT2 are in much closer agreement with the experimental data than RAS(SD)-1SF. 
This difference in performance of these methods can be attributed to the importance of dynamical electron correlation that is largely missing in RAS(SD)-1SF, but is incorporated in QDNEVPT2 up to the second order in multireference perturbation theory.

\subsection{Ground- and excited-state spin--orbit coupling in $3d$ and $4d$ transition metal ions}
\label{sec:results:3d_4d}

\begin{figure*}[t!]
	\subfigure[]{\includegraphics[width=0.48\textwidth]{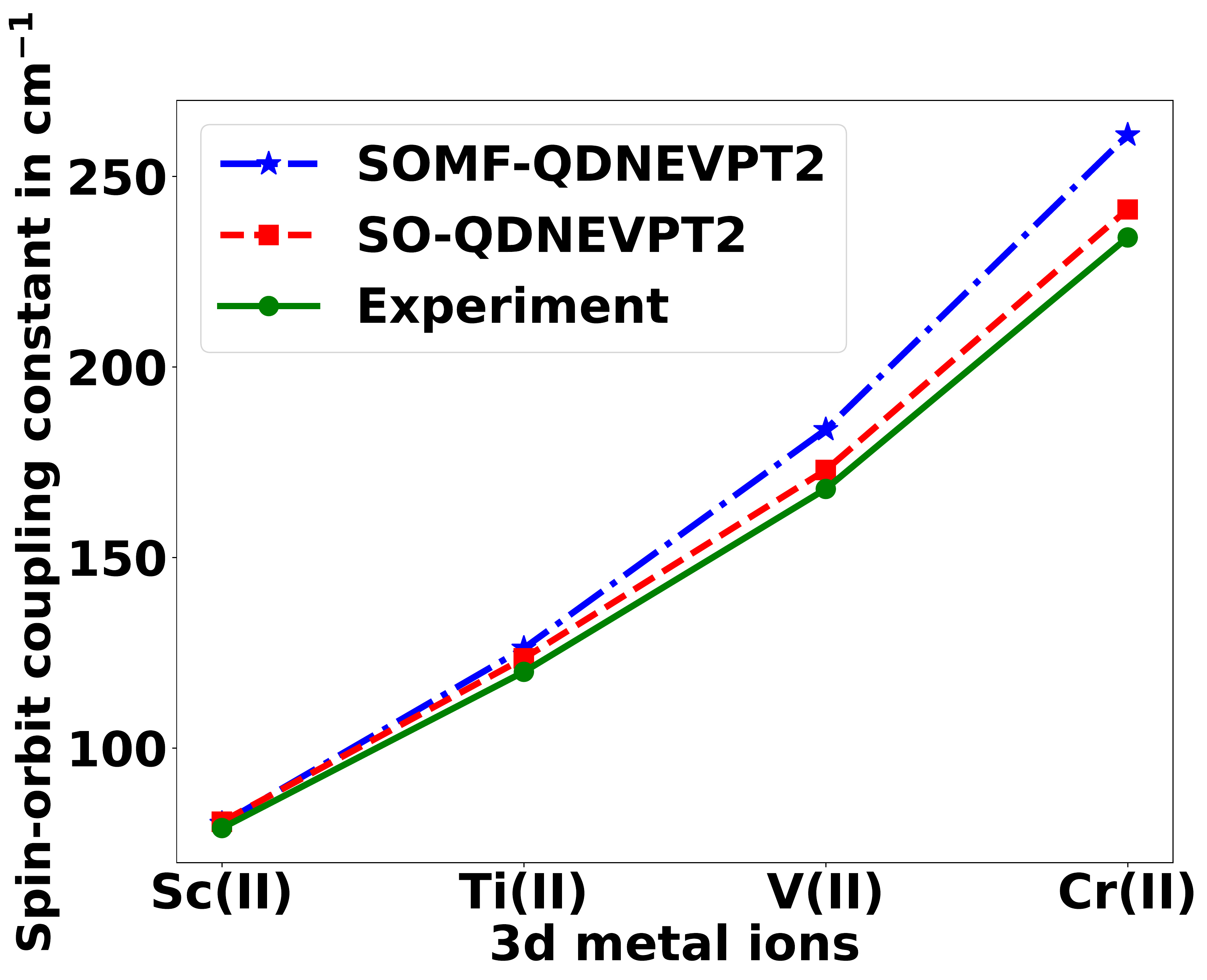} \label{fig:gs_socc_3d_1}} 
	\subfigure[]{\includegraphics[width=0.48\textwidth]{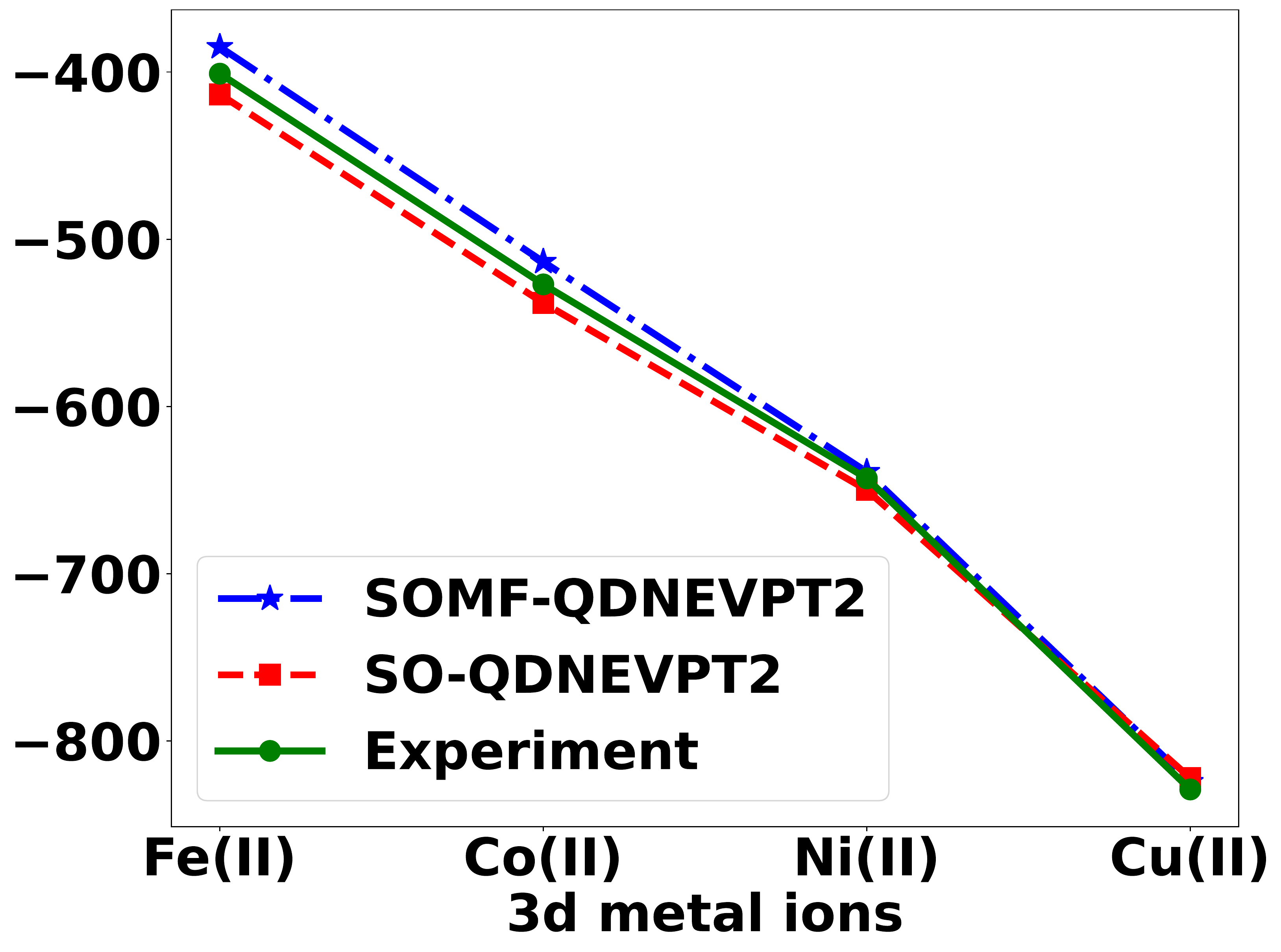} \label{fig:gs_socc_3d_2}}  
	\subfigure[]{\includegraphics[width=0.48\textwidth]{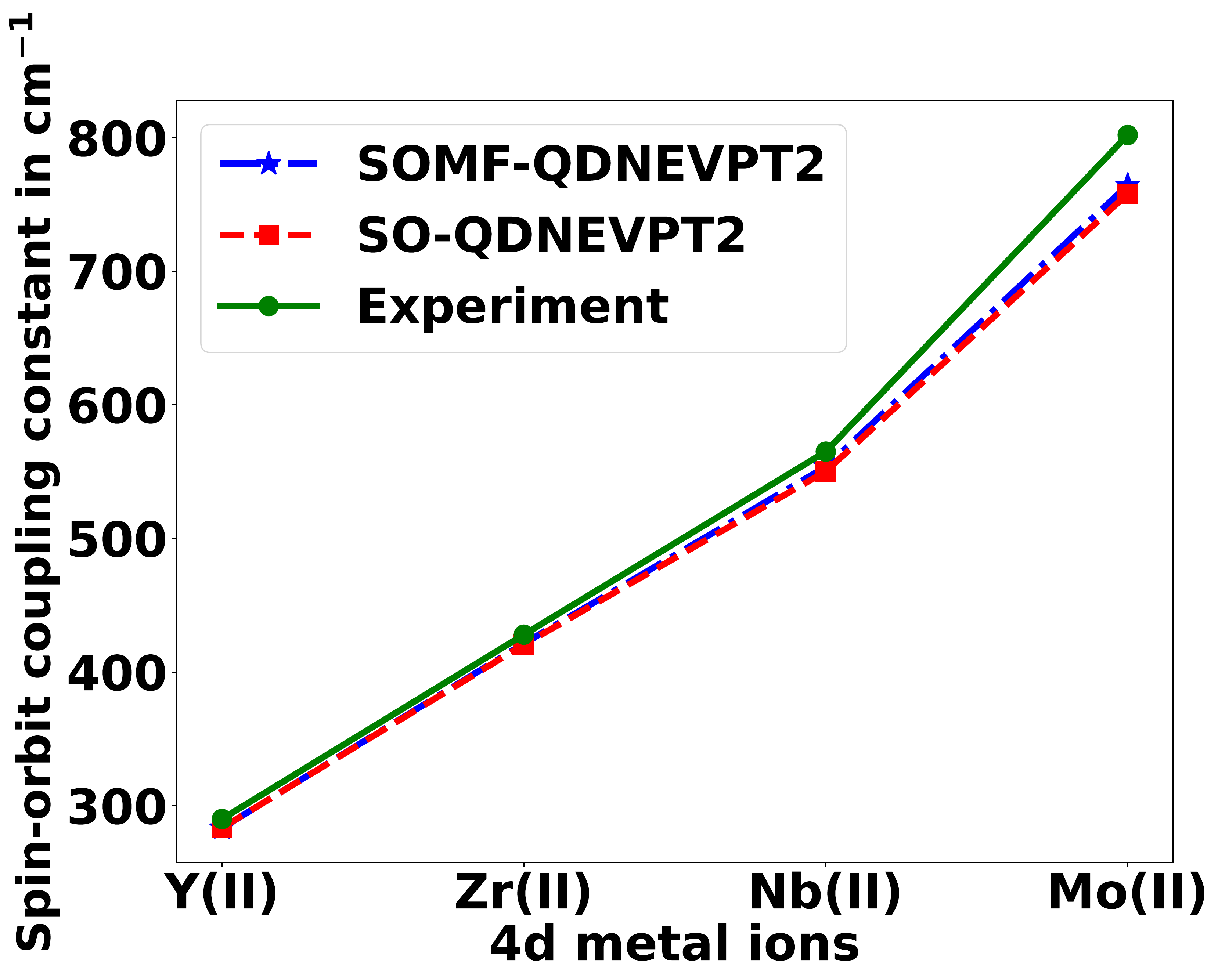} \label{fig:gs_socc_4d_1}} 
	\subfigure[]{\includegraphics[width=0.48\textwidth]{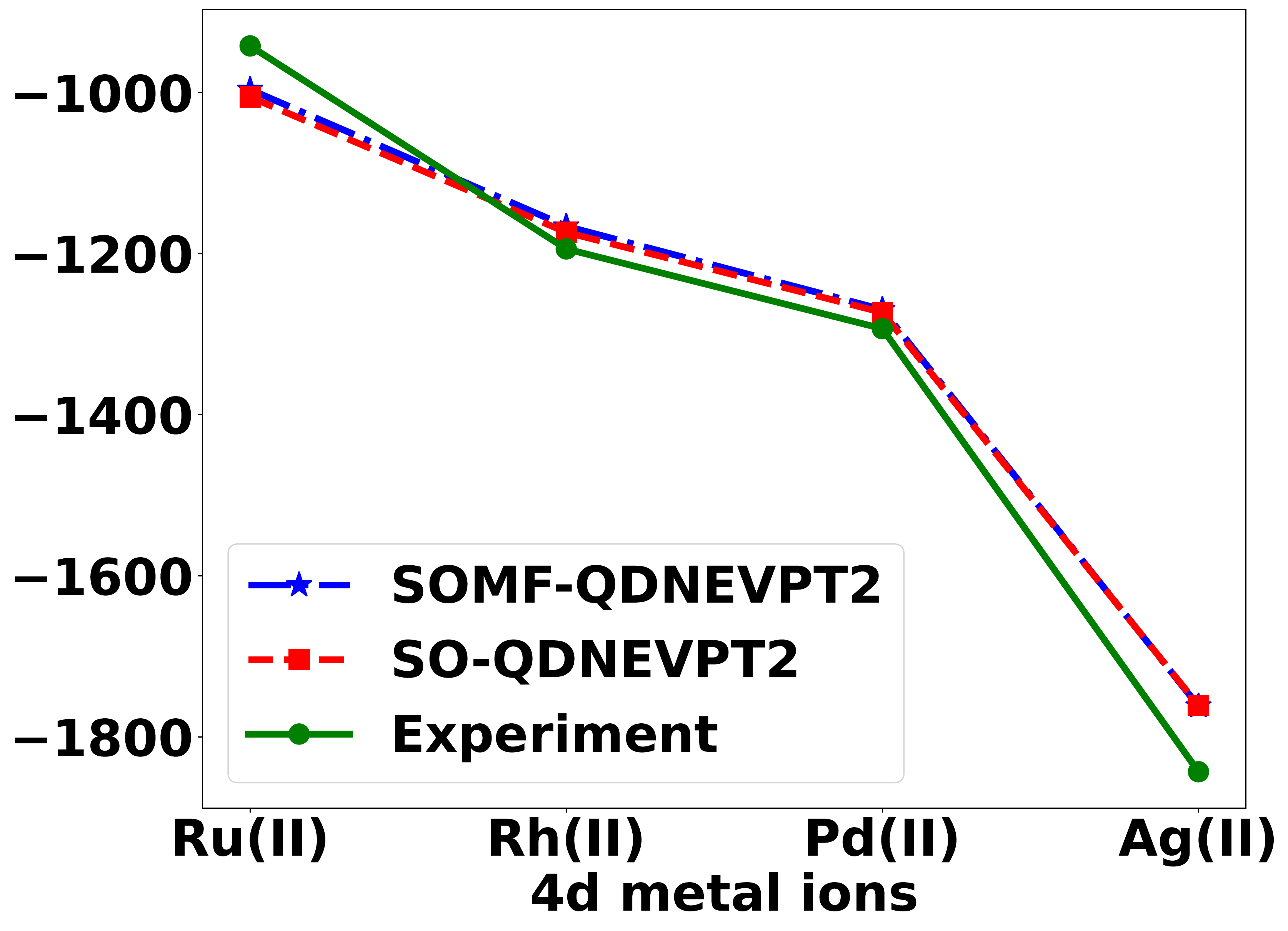} \label{fig:gs_socc_4d_2}}  
	\captionsetup{justification=raggedright,singlelinecheck=false}
	\caption{Total spin--orbit coupling constants (\cm) calculated for the ground electronic terms of $3d$ (a, b) and $4d$ (c, d) transition metal ions (M$^{2+}$) using SO- and SOMF-QDNEVPT2 in comparison to experimental data.\cite{Epstein:1975p310,Sugar:1985,Forbes:1997p310,Gayazov:1998p45,Sugar:2009p527,Smillie:2016p12,NIST_ASD}
	}
	\label{fig:gs_socc}
\end{figure*}

To assess the performance of SO-QDNEVPT2 and SOMF-QDNEVPT2 for transition metal systems, we calculated ZFS in the ground and excited states of $3d$ and $4d$ metal ions with the 2+ charge (M$^{2+}$). 
We consider all M$^{2+}$ ions with electronic configurations $nd^1$ to $nd^9$ except $nd^5$, which does not show spin--orbit coupling in the ground $^6S$ state.
In the weak LS-coupling regime, the energy levels of spin--orbit-coupled states $E_J$ can be expressed as follows:\cite{Condon:1951}
\begin{align}
	E_J 
	&= E_{LS} \notag \\
	&+ \frac{1}{2}\lambda[J(J+1) - L(L+1) -S(S+1)] \ ,
\end{align}
where $E_{LS}$ is the energy of electronic term with quantum numbers $L$ and $S$ that does not incorporate spin--orbit coupling, $J$ is the quantum number of total angular momentum, and $\lambda$ is the spin--orbit coupling constant (SOCC), which is related to the energy spacing between two levels:
\begin{align}
	\label{eq:socc}
	E_J - E_{J-1} = \lambda J  \ .
\end{align}
Since $E_J$ increases with increasing $J$ for $nd^1$ to $nd^4$ and decreases with increasing $J$ for $nd^6$ to $nd^9$, $\lambda$ can take either positive or negative values. 
In practice, the SOCC calculated using \cref{eq:socc} for a particular electronic term show dependence on $J$ and have different values for different pairs of energy levels $E_J$ and $E_{J-1}$.
To quantify ZFS in M$^{2+}$ using a single parameter, we compute the total SOCC
\begin{align}
	\label{eq:total_socc}
	\Lambda = \sum_J \lambda_J
\end{align}
where  $\lambda_J$ is obtained using \cref{eq:socc}.

\begin{figure*}[t!]
	\subfigure[]{\includegraphics[width=0.48\textwidth]{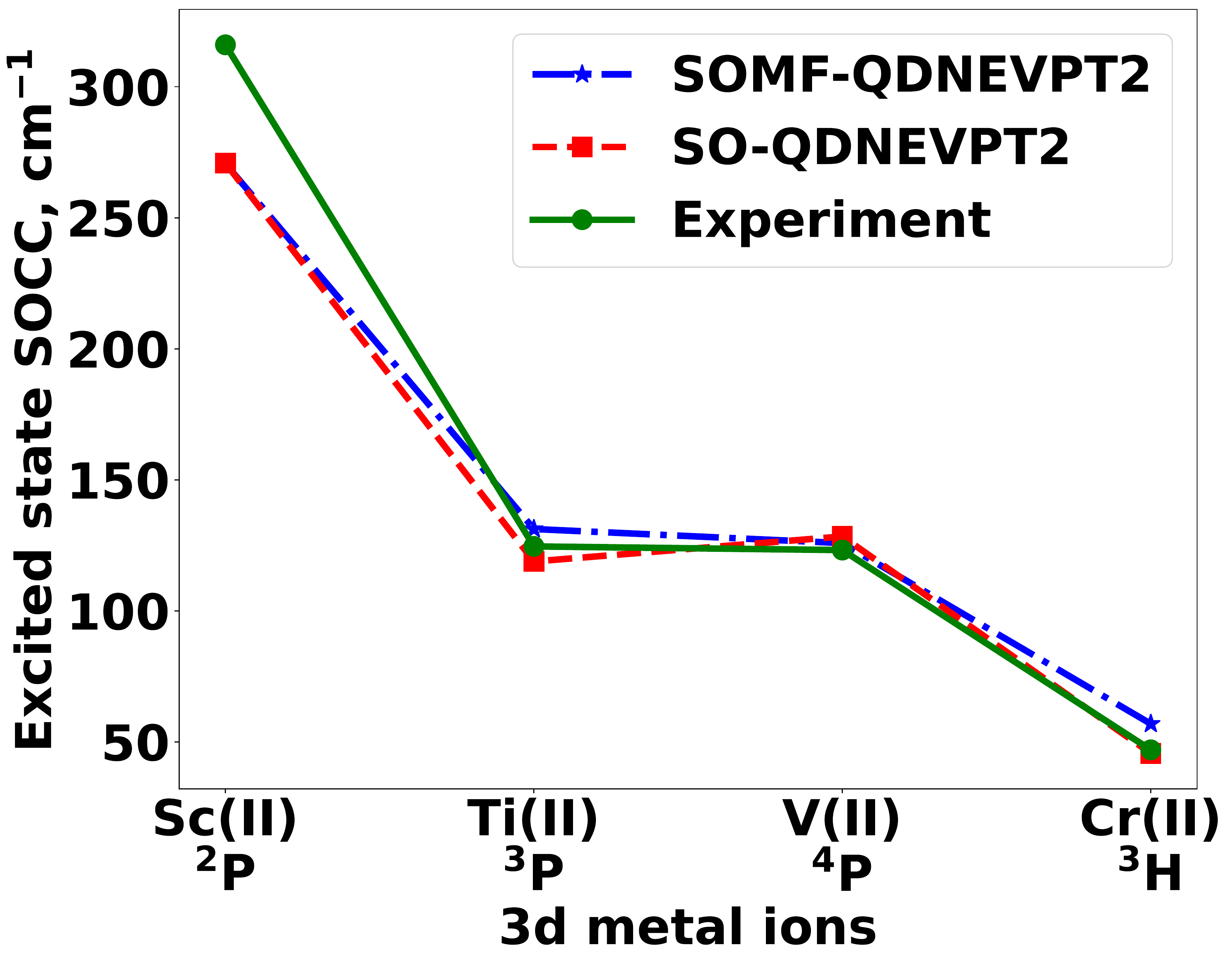} \label{fig:es_socc_3d_1}} 
	\subfigure[]{\includegraphics[width=0.48\textwidth]{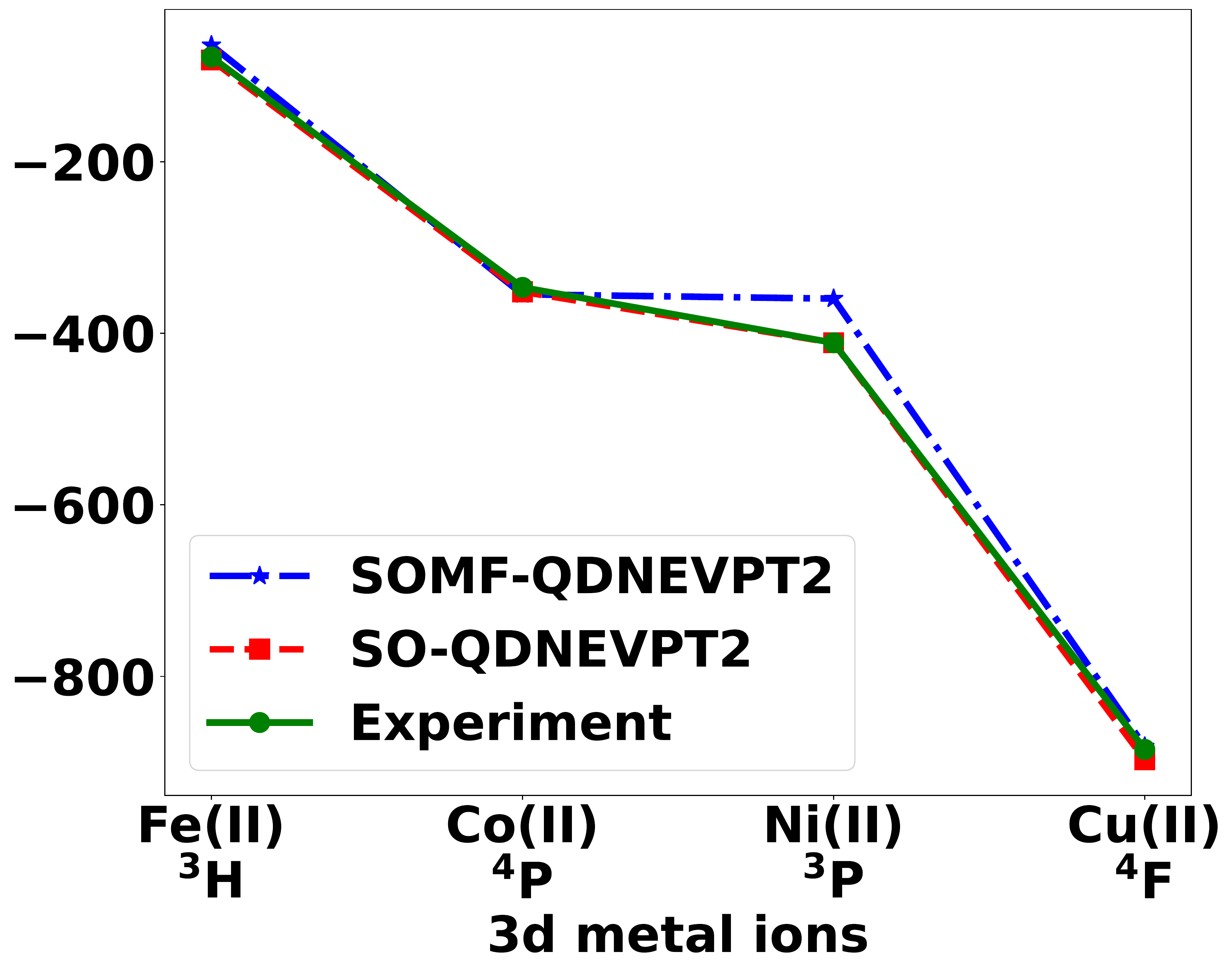} \label{fig:es_socc_3d_2}}
	\subfigure[]{\includegraphics[width=0.48\textwidth]{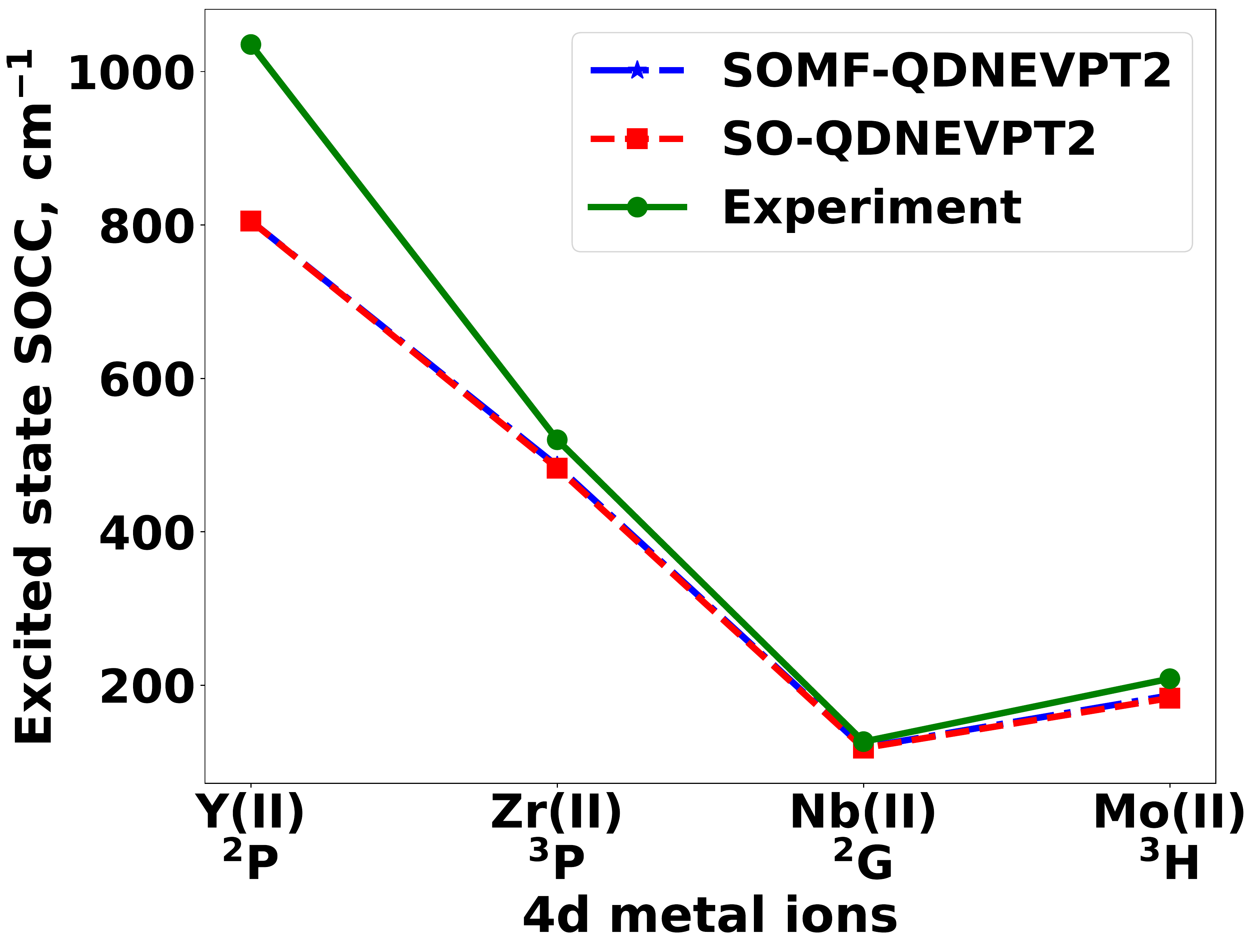} \label{fig:es_socc_4d_1}} 
	\subfigure[]{\includegraphics[width=0.48\textwidth]{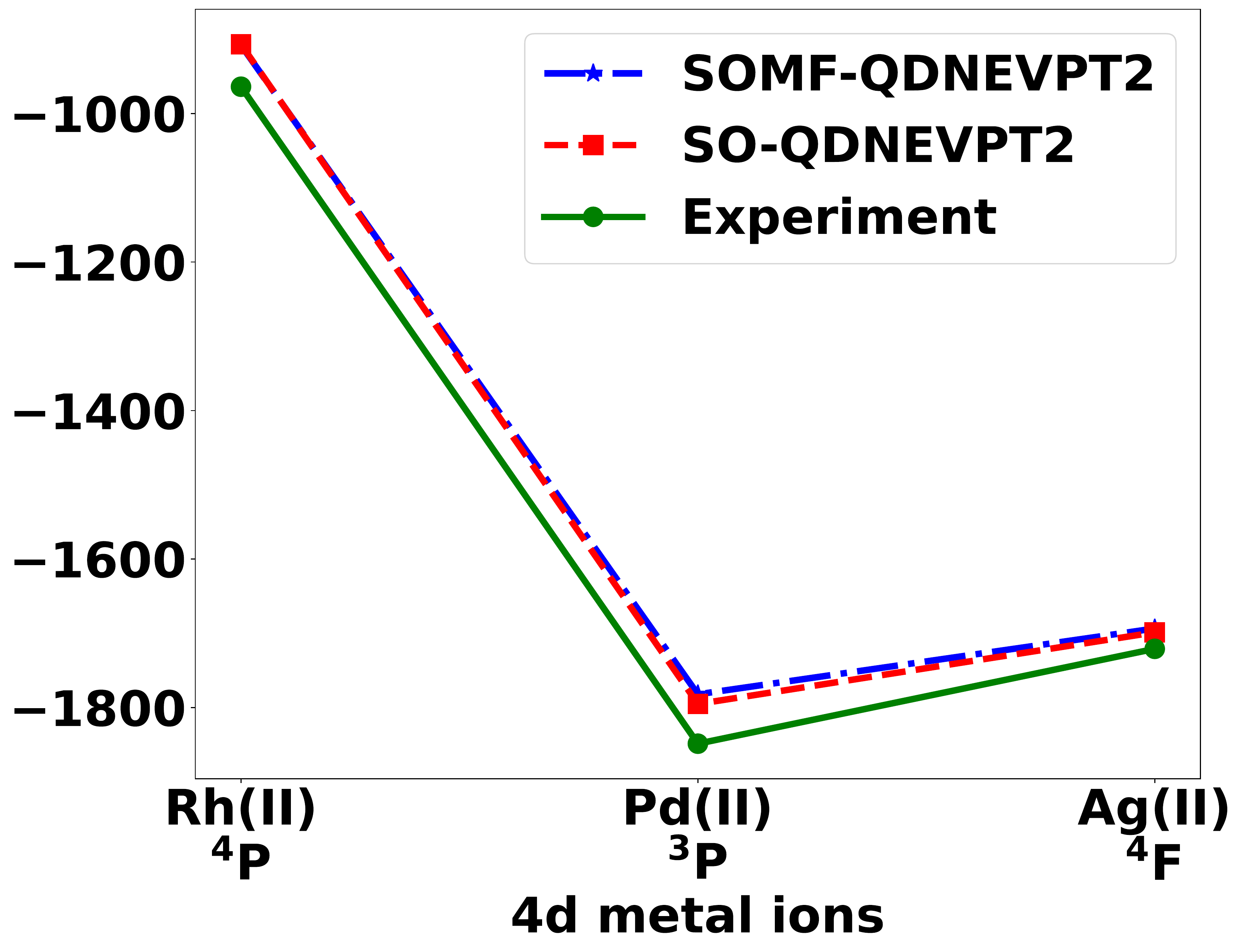} \label{fig:es_socc_4d_2}}
	\captionsetup{justification=raggedright,singlelinecheck=false}
	\caption{
		Total spin--orbit coupling constants (\cm) calculated for the excited electronic terms of $3d$ (a, b) and $4d$ (c, d) transition metal ions (M$^{2+}$) using SO- and SOMF-QDNEVPT2 relative to experimental data.\cite{Epstein:1975p310,Sugar:1985,Forbes:1997p310,Gayazov:1998p45,Sugar:2009p527,Smillie:2016p12,NIST_ASD}
	}
	\label{fig:es_socc}
\end{figure*}

\cref{fig:gs_socc} shows the total SOCC ($\Lambda$) calculated using the QDNEVPT2 methods and experimental data for the ground electronic terms of $3d$ and $4d$ transition metal ions, respectively.
In each row of periodic table, the magnitude of $\Lambda$ increases with increasing nuclear charge. 
For the $3d$ metal ions, the SO-QDNEVPT2 and SOMF-QDNEVPT2 results show significant differences (\cref{fig:gs_socc_3d_1,fig:gs_socc_3d_2}).
The best agreement with the experiment\cite{NIST_ASD} is shown by SO-QDNEVPT2 that predicts $\Lambda$ with errors of 3.1 \% or less. 
The SOMF-QDNEVPT2 method yields larger $\Lambda$ overestimating the experimental SOCC by up to 11.5 \%. 
The most noticeable errors of SOMF approximation are observed in the middle of $3d$ transition metal row (V$^{2+}$, Cr$^{2+}$, Fe$^{2+}$, and Co$^{2+}$), indicating that the two-electron spin--orbit interactions neglected in SOMF are important for these metal ions.  
In contrast to the $3d$ ions, for the $4d$ transition metal row SO-QDNEVPT2 and SOMF-QDNEVPT2 predict very similar SOCC that differ by less than 10 \cm ($<$ 1 \%) from each other (\cref{fig:gs_socc_4d_1,fig:gs_socc_4d_2}).
When compared to the experimental data, the errors of QDNEVPT2 methods in $4d$ SOCC do not exceed 6.7 \%. 
The higher accuracy of SOMF approximation in the $4d$ metal ions may be attributed to the greater radial extent of $4d$ orbitals compared to that in $3d$ orbitals leading to a reduced contribution from two-electron spin--orbit coupling effects.

\cref{fig:es_socc} shows the SO-QDNEVPT2 and SOMF-QDNEVPT2 errors in total SOCC for the selected excited electronic terms of $3d$ and $4d$ metal ions. 
In these calculations, we excluded Ru$^{2+}$, which exhibited convergence problems when excited electronic states were included in SA-CASSCF.
As in \cref{fig:gs_socc}, SOMF-QDNEVPT2 shows significantly larger SOMF errors in the excited-state $\Lambda$ of $3d$ metal ions compared to those of $4d$ ions (\cref{fig:es_socc_3d_1,fig:es_socc_3d_2}). 
These errors of SOMF approximation become particularly noticeable for the ions with two (or more) electrons or holes in the $d$-shell (Ti to Ni) where they contribute up to 25 \% of the total SOMF-QDNEVPT2 error in SOCC.
For the excited states of $4d$ metal ions, the SOMF approximation is once again very accurate, resulting in similar SOCC computed using SO-QDNEVPT2 and SOMF-QDNEVPT2 (\cref{fig:es_socc_4d_1,fig:es_socc_4d_2}).
Overall, the best agreement with experimental data is demonstrated by SO-QDNEVPT2 that is significantly more accurate than SOMF-QDNEVPT2 for the $3d$ metals ions.

\subsection{Low-lying electronic states of \ce{NpO2^2+} and \ce{PuO2^2+}}
\label{sec:results:Np_Pu}

	\begin{table*}[t!]
	\caption{Excited-state energies (in \cm) of \ce{NpO2^2+} computed using four methods, relative to the ${}^2\Phi_{5/2u}$ ground state. The QDNEVPT2 and CASPT2-SO\cite{Gendron:2014p8577} methods employed the (7e, 10o) active space and the ANO-RCC-VTZP basis set. In the SO-SHCI calculations,\cite{Mussard:2018p154} the (17e, 143o) active space was used. 
	} 
	\label{tab:npo}
	\setstretch{1}
	\small
	\centering
	\begin{threeparttable}
		\begin{tabular}{c c c c c}
			\hline\hline
			Electronic state   		& SOMF-QDNEVPT2 & SO-QDNEVPT2 & CASPT2-SO\cite{Gendron:2014p8577} & SO-SHCI\tnote{a} \\ \hline
			${}^2\Phi_{5/2u}$  	& 0.0               & 0.0            & 0.0                           & 0.0              \\
			${}^2\Delta_{3/2u}$   & 3549.2        & 3550.7      & 3107                        & 3857             \\
			${}^2\Phi_{7/2u}$  	& 8000.4        & 8001.1      & 8080                        & 8675             \\
			${}^2\Delta_{5/2u}$   & 9470.4        & 9470.2      & 9313                        & 10077            \\ \hline\hline
		\end{tabular}
		\begin{tablenotes}
			\item[a] The SO-SHCI excitation energies from Ref.\@ \citenum{Mussard:2018p154} used a modified ANO-RCC-VTZP basis set with the $5s4p2d1f$ contraction for the oxygen atoms.
		\end{tablenotes}			
	\end{threeparttable}
\end{table*}

\begin{table*}[t!]
	\caption{Contributions (in \%) to the spin--orbit-coupled electronic states of \ce{NpO2^2+} computed using SO-QDNEVPT2 and CASPT2-SO\cite{Gendron:2014p8577} methods. } 
	\label{tab:npo_state}
	\setstretch{1}
	\small
	\centering
	\begin{threeparttable}
		\begin{tabular}{c c c c c}
			\hline\hline
			Electronic state   & SO-QDNEVPT2                     & CASPT2-SO\cite{Gendron:2014p8577} &  \\ \hline
			${}^2\Phi_{5/2u}$  & 89.1 $^2\Phi_u$ + 10.6 $^2\Delta_u$ & 88 $^2\Phi_u$ + 12 $^2\Delta_u$ &  \\
			${}^2\Delta_{3/2u}$ & 98.5 $^2\Delta_u$ + 1.4 $^2\Pi_u$   & 98 $^2\Delta_u$ + 2 $^2\Pi_u$   &  \\
			${}^2\Phi_{7/2u}$  & 99.8 $^2\Phi_u$                   & 100 $^2\Phi_u$                &  \\
			${}^2\Delta_{5/2u}$ & 89.4 $^2\Delta_u$ + 10.5 $^2\Phi_u$ & 89 $^2\Delta_u$ + 11 $^2\Phi_u$ &  \\ \hline\hline
		\end{tabular}
	\end{threeparttable}
\end{table*}

Finally, to test the limits of SO-QDNEVPT2 and SOMF-QDNEVPT2 applicability, we use these methods to compute the low-lying electronic states of two actinide dioxides, neptunyl (VI) (\ce{NpO2^2+}) and plutonyl (VI) (\ce{PuO2^2+}) dications, which present major challenges for theories that employ perturbative treatment of spin--orbit coupling.\cite{Fujii:2015p1015,Pegg:2019p760,Gendron:2014p8577,Gendron:2014p7994,Mussard:2018p154} 

In \ce{NpO2^2+}, the spin--orbit coupling mixes the $^2\Phi_u$ and $^2\Delta_u$ electronic terms originating from $5f^1$ configuration, which gives rise to the ${}^2\Phi_{5/2u}$, ${}^2\Delta_{3/2u}$, ${}^2\Phi_{7/2u}$, and ${}^2\Delta_{5/2u}$ electronic states.
The relative energies of these states computed using SO-QDNEVPT2 and SOMF-QDNEVPT2 are presented in \cref{tab:npo}.
For comparison, we also show the results from the CASPT2-SO study by Gendron et al.\@ that employs the perturbative treatment of spin--orbit coupling using the Douglas--Kroll--Hess (DKH) Hamiltonian\cite{Gendron:2014p8577} and from the variational implementation of spin--orbit semistochastic heat bath configuration interaction (SO-SHCI) by Mussard et al.\@ employing the two-component X2C Hamiltonian.\cite{Mussard:2018p154}
All excitation energies reported in \cref{tab:npo} were calculated using the same molecular geometry and the ANO-RCC-VTZP basis set (180 molecular orbitals), with the exception of SO-SHCI calculations where ANO-RCC-VTZP was modified by including eight additional basis functions as described in \cref{tab:npo} (188 molecular orbitals). 
Since the SO-SHCI calculations achieved the highest level of electron correlation and spin--orbit coupling treatment in the (17e, 143o) active space, we consider their results as the theoretical best estimate of excitation energies in \ce{NpO2^2+}.
We note, however, that the SO-SHCI study did not incorporate dynamical correlation for the 90 electrons outside the active space, which was accounted for in the SO-QDNEVPT2, SOMF-QDNEVPT2, and CASPT2-SO calculations.

The best agreement with SO-SHCI in \cref{tab:npo} is shown by SO-QDNEVPT2 and SOMF-QDNEVPT2, which predict the ${}^2\Delta_{3/2u}$, ${}^2\Phi_{7/2u}$, and ${}^2\Delta_{5/2u}$ excitation energies with the mean absolute error (MAE) of $\sim$ 529 \cm. 
Due to the one-electron character of all excitations in \ce{NpO2^2+}, the errors introduced by the SOMF approximation are less than 2 \cm.
The CASPT2-SO method exhibits larger errors for the ${}^2\Delta_{3/2u}$ and ${}^2\Delta_{5/2u}$ states and MAE of 703 \cm relative to SO-SHCI.
\cref{tab:npo_state} demonstrates that both types of multireference perturbation theories predict similar composition of spin--orbit-coupled electronic states, estimating the mixing between $^2\Phi_u$ and $^2\Delta_u$ for $J$ = 5/2 of $\sim$ 11 to 12 \%.

\begin{table*}[t!]
	\caption{Excited-state energies (in \cm) of \ce{PuO2^2+} computed using three methods and the ANO-RCC-VTZP basis set, relative to the $4_g$ ground state. The QDNEVPT2 and CASPT2-SO\cite{Gendron:2014p8577} methods employed the (8e, 10o) active space.} 
	\label{tab:puo}
	\setstretch{1}
	\small
	\begin{threeparttable}
		\begin{tabular}{c c c c } 
			\hline\hline
			Electronic state  & SOMF-QDNEVPT2 & SO-QDNEVPT2 & CASPT2-SO\cite{Gendron:2014p8577}  \\
			\hline
			$4_g$                 	& 0.0 & 0.0 & 0.0\\ 
			$0^{+}_g$   		& 2924.9 &  2922.3 & 3132 \\
			$1_g$               	& 5176.5&  5169.0 & 5464\\
			$5_g$  				& 7197.2 & 7186.9 & 7238\\
			$0^{-}_g$ 			& 10679.0& 10673.7 & 11171\\ 
			$1_g$        			& 11393.1 & 11375.0 & 11682 \\
			\hline\hline
		\end{tabular}
	\end{threeparttable}
\end{table*}

\begin{table*}[t!]
	\caption{Contributions (in \%) to the spin--orbit-coupled electronic states of \ce{PuO2^2+} computed using SO-QDNEVPT2 and CASPT2-SO\cite{Gendron:2014p8577} methods. } 
	\label{tab:puo_state}
	\setstretch{1}
	\small
	\begin{threeparttable}
		\begin{tabular}{c c c } 
			\hline \hline
			Electronic state  & SO-QDNEVPT2 & CASPT2-SO\cite{Gendron:2014p8577} \\
			\hline
			$4_g$                 & 95.4 ${}^3H_g$ + 3.8 ${}^1\Gamma_g$ & 98 ${}^3H_g$ + 2 ${}^1\Gamma_g$\\ 
			$0^{+}_g$   & 53.4 ${}^3\Sigma^{-}_g$ + 30.7 ${}^3\Pi_g$ + 14.0 ${}^1\Sigma^{+}_g$ &  54 ${}^3\Sigma^{-}_g$ + 26 ${}^3\Pi_g$ + 17 ${}^1\Sigma^{+}_g$ \\
			$1_g$           & 52.6 ${}^3\Pi_g$ + 25.9 ${}^3\Sigma^{-}_g$ + 18.8 ${}^1\Pi_g$  & 49 ${}^3\Pi_g$ + 26 ${}^3\Sigma^{-}_g$ + 23 ${}^1\Pi_g$\\
			$5_g$  & 98.8 ${}^3H_g$ & 99 ${}^3H_g$\\
			$0^{-}_g$ & 99.9 ${}^3\Pi_g$ & 100 ${}^3\Pi_g$ \\ 
			$1_g$        & 69.9 ${}^3\Sigma^{-}_g$ + 19.4 ${}^1\Pi_g$ + 8.5 ${}^3\Pi_g$ & 70 ${}^3\Sigma^{-}_g$ + 17 ${}^1\Pi_g$ + 8 ${}^3\Pi_g$ \\
			\hline\hline
		\end{tabular}
	\end{threeparttable}
\end{table*}

The excited-state energies of \ce{PuO2^2+} computed using SO-QDNEVPT2, SOMF-QDNEVPT2, and CASPT2-SO\cite{Gendron:2014p8577} are shown in \cref{tab:puo}. 
Due to the $5f^2$ configuration of Pu, the energy level diagram of \ce{PuO2^2+} is much more complicated than that of  \ce{NpO2^2+} with several electronic terms mixing with each other upon incorporating the spin--orbit coupling effects.
The SO-QDNEVPT2 and CASPT2-SO calculations show similar results.
Both methods predict the same ordering of electronic states with excitation energies differing by less than 500 \cm.
As shown in \cref{tab:puo_state}, SO-QDNEVPT2 and CASPT2-SO also agree in the assignments of each state, predicting the contributions from each electronic term within 5\% of each other.
Introducing the SOMF approximation changes the excitation energies by at most 18.1 \cm, which is noticeably greater than the SOMF error in \ce{NpO2^2+}, but is much smaller than the energy spacing between spin--orbit-coupled states.

\section{Conclusions}
\label{sec:conclusions}

In this work, we presented the first implementation of spin--orbit coupling effects in fully internally contracted second-order quasidegenerate $N$-electron valence perturbation theory (QDNEVPT2).
Our implementation provides two methods for incorporating spin--orbit coupling up to the first order in perturbation theory: 1) using the full Breit--Pauli (BP) relativistic Hamiltonian (SO-QDNEVPT2) and 2) approximating the BP Hamiltonian using the spin--orbit mean-field approach (SOMF-QDNEVPT2).
The SO-QDNEVPT2 and SOMF-QDNEVPT2 methods have several attractive features: 
i) they combine the description of static electron correlation with a computationally efficient treatment of dynamic correlation and spin--orbit coupling in near-degenerate electronic states; 
ii) they are fully invariant with respect to the transformations within the subspaces of core, active, and external molecular orbitals;
iii) they achieve a lower computational scaling with the active space size than conventional QDNEVPT2 by avoiding the calculation of four-particle reduced density matrices without introducing any approximations;
iv) they take advantage of full internal contraction while preserving the degeneracy of spin--orbit-coupled states;
and v) they enable computing transition properties, such as oscillator strengths.
In addition, comparing the results of SO-QDNEVPT2 and SOMF-QDNEVPT2 allows to quantify and systematically analyze the errors of SOMF approximation.

To demonstrate the capabilities of SO-QDNEVPT2 and SOMF-QDNEVPT2 and benchmark their accuracy, we computed the zero-field splitting (ZFS) in the ground electronic states of group 14 and 16 hydrides, the ground and excited states of $3d$ and $4d$ transition metal ions, and the low-lying electronic states of actinide oxides (\ce{NpO2^2+} and \ce{PuO2^2+}).
Our results demonstrate that SO-QDNEVPT2 predicts accurate ZFS for the compounds of elements up to the fourth row of periodic table where errors of less than 5 \% relative to experimental data are observed. 
For the fifth-row elements (in SnH, TeH, and $4d$ transition metal ions), the errors in ZFS increase up to $\sim$ 10 \%.
In actinides, the SO-QDNEVPT2 results are in a good agreement with the data from CASPT2-SO and SO-SHCI methods for the energy spacings between electronic states and the characters of their wavefunctions.
The SOMF-QDNEVPT2 and SO-QDNEVPT2 results are very similar to each other for all systems but the $3d$ transition metal ions, where the SOMF approximation significantly increases the errors in computed ZFS relative to experiment.

Overall, our results demonstrate that SO-QDNEVPT2 and SOMF-QDNEVPT2 are promising approaches for simulating spin--orbit coupling in the ground and excited states of chemical systems with multireference electronic structure. 
Future work in our group will focus on improving the accuracy of these methods for the heavier ($>$4th row) elements and their extensions to simulate the magnetic properties of molecules. 

\suppinfo
See Supplementary Information for the numerical assessment of amplitude averaging in \cref{eq:mod_state_averaging}, composition of active spaces in the calculations of group 14 and 16 hydrides, additional computational details for the study of $3d$ and $4d$ transition metal ions, and the results of SA-CASSCF and QDNEVPT2 calculations for \ce{NpO2^2+} and \ce{PuO2^2+}.

\acknowledgement
This work was supported by the start-up funds from the Ohio State University. 
Additionally, A.Y.S. was supported by National Science Foundation, under Grant No.\@ CHE-2044648.
The authors would like to thank Lan Cheng, Xubo Wang, and Sandeep Sharma for insightful discussions.


\providecommand{\latin}[1]{#1}
\makeatletter
\providecommand{\doi}
  {\begingroup\let\do\@makeother\dospecials
  \catcode`\{=1 \catcode`\}=2 \doi@aux}
\providecommand{\doi@aux}[1]{\endgroup\texttt{#1}}
\makeatother
\providecommand*\mcitethebibliography{\thebibliography}
\csname @ifundefined\endcsname{endmcitethebibliography}
  {\let\endmcitethebibliography\endthebibliography}{}

\end{document}